\documentclass{article}

\PassOptionsToPackage{round,authoryear}{natbib}

\usepackage[preprint]{neurips_2024}

\usepackage[utf8]{inputenc} 
\usepackage[T1]{fontenc}    
\usepackage{hyperref}       
\usepackage{url}            
\usepackage{booktabs}       
\usepackage{amsfonts}       
\usepackage{nicefrac}       
\usepackage{microtype}      
\usepackage{xcolor}         
\usepackage{graphicx}
\usepackage{subcaption}
\usepackage{multirow}
\usepackage{amsmath}
\usepackage{placeins}

\usepackage{xcolor}
\definecolor{lowc}{RGB}{0, 128, 255}
\definecolor{mediumc}{RGB}{255, 179, 0}
\definecolor{highc}{RGB}{229, 99, 153}

\usepackage{pifont}

\usepackage{amsthm}
\theoremstyle{definition}
\newtheorem{definition}{Definition}[]

\title{Towards Modeling Road Access Deprivation in Sub-Saharan Africa Based on a New Accessibility Metric and Road Quality}

\author{%
Sebastian Hafner$^{1}$ \quad Qunshan Zhao$^{1*}$ \quad Bunmi Alugbin$^2$ \quad Kehinde Baruwa$^2$ \\ \textbf{Caleb Cheruiyot}$^3$ \quad \textbf{Sabitu Sa'adu Da'u}$^4$ \quad \textbf{Xingyi Du}$^{1}$ \quad \textbf{Peter Elias}$^2$ \\ \textbf{Helen Elsey}$^5$ \quad \textbf{Ryan Engstrom}$^6$ \quad \textbf{Serkan Girgin}$^7$ \quad \textbf{Diego F. P. Grajales}$^{1}$ \\ \textbf{Esther Judith}$^3$ \quad \textbf{Caroline Kabaria}$^3$ \quad \textbf{Monika Kuffer}$^7$ \quad \textbf{Oluwatoyin Odulana}$^2$ \\ \textbf{Francis C. Onyambu}$^3$ \quad \textbf{Adenike Shonowo}$^{1}$ \quad \textbf{Dana R. Thomson}$^8$ \quad \textbf{Mingyu Zhu}$^1$ \\ \textbf{João Porto de Albuquerque}$^1$ \\
$^1$University of Glasgow \quad $^2$University of Lagos \quad $^3$APHRC \quad $^4$Bayero University \\ $^5$University of York \quad $^6$George Washington University \quad $^7$University of Twente \\
$^8$Columbia University \\
$^*$Corresponding author: \texttt{qunshan.zhao@glasgow.ac.uk}
}

\begin{document}

\maketitle

\begin{abstract}
Access to motorable roads is a critical dimension of urban infrastructure, particularly in rapidly urbanizing regions such as Sub-Saharan Africa. Yet, many urban communities, especially those in informal settlements, remain disconnected from road networks. This study presents a road access deprivation model that combines a new accessibility metric, capturing how well buildings are connected to the road network, with road surface type data as a proxy for road quality. These two components together enable the classification of urban areas into low, medium, or high deprivation levels. The model was applied to Nairobi (Kenya), Lagos (Nigeria), and Kano (Nigeria) using open geospatial datasets. Across all three cities, the majority of built-up areas fall into the low and medium road access deprivation levels, while highly deprived areas are comparatively limited. However, the share of highly deprived areas varies substantially, ranging from only 11.8 \% in Nairobi to 27.7 \% in Kano. Model evaluation against community-sourced validation data indicates good performance for identifying low deprivation areas (F1 $>$ 0.74), moderate accuracy for medium deprivation in Nairobi and Lagos (F1 $>$ 0.52, lower in Kano), and more variable results for high deprivation (F1 ranging from 0.26 in Kano to 0.69 in Nairobi). Furthermore, analysis of grid cells with multiple validations showed strong agreement among community members, with disagreements occurring mainly between adjacent deprivation levels. Finally, we discussed two types of sources for disagreement with community validations: (1) misalignment between the conceptual model and community perceptions, and (2) the operationalization of the conceptual model. In summary, our road access deprivation modeling approach demonstrates promise as a scalable, interpretable tool for identifying disconnected areas and informing urban planning in data-scarce contexts.
\end{abstract}

\section{Introduction}


Urban access to roads is a critical dimension of development and a key enabler for service delivery, economic participation, and spatial integration \citep{unhabitat2014streets}. In rapidly urbanizing regions such as Sub-Saharan Africa, road networks serve as a backbone for mobility, emergency response, and infrastructure provision. However, large segments of urban populations, especially those residing in informal settlements, remain inadequately connected to motorable roads, reinforcing spatial inequality and limiting development opportunities \citep{bettencourt2025infrastructure}. As a consequence, for example, fires are very recurrent without the option of firetrucks accessing densely built-up settlements.

A growing body of literature has emphasized the importance of well-planned street networks for urban poverty reduction. \citet{angel2011making} and \citet{unhabitat2014streets} argue that coherent road layouts support inclusive growth by improving accessibility and service reach. More recently, the Million Neighborhoods project developed a topological analysis to quantify the relationship between roads and buildings \citep{brelsford2018toward, brelsford2019optimal}. \cite{brelsford2018toward} introduced block complexity, a measurement of the connectedness of a city block. A higher complexity value is associated with increased difficulty in reaching places within the block from the street network. They demonstrated that this topological analysis is effective in identifying slums based on the lack of spatial access and related services, using a township of Cape Town, South Africa, as an example. In \cite{brelsford2019optimal}, they provide further evidence of the effectiveness of their topological analysis in revealing the absence of access to places within a slum. Beyond that, they proposed a method to reduce block complexity via topological optimization, generating an access network that makes each structure in a settlement accessible. 

Enabled by the availability of open building footprint and street network data, topological analysis of street access can now be performed at large scales \citep{soman2020worldwide, bettencourt2025infrastructure, thomson2025city}. \cite{soman2020worldwide} applied the concept of block complexity to the entire Global South using OpenStreetMap (OSM) data. More recently, \cite{bettencourt2025infrastructure} combined building footprints data from Ecopia Landbase Africa and OSM street network data to compute block complexity in Sub-Saharan Africa. They demonstrated that block complexity expresses not only spatial access deprivation but also entails many dimensions of low human development. For example, they showed that higher block complexity is systematically associated with lower female literacy, lower access to a water source on premises, or lower access to an improved sanitation facility. In contrast, \cite{thomson2025city} developed the city segment layer, which identifies sub-city areas characterized by high building density and low road connectivity, serving as a proxy for neighborhood socioeconomic status. Unlike block complexity, the city segment layer measures road connectivity in terms of meters of road per building.

Beyond street access, studies have demonstrated that road surface type serves as a useful proxy for identifying informal or underserved urban areas \citep{yeboah2021analysis, kohli2012ontology}. The importance of road surface type was further highlighted in a recent study that found a positive correlation between average road pavedness and human development index values across different countries \citep{randhawa2025paved}. In particular, significant gaps in paved road coverage were noted in specific areas of Africa \citep{randhawa2025paved}. Road quality has also been linked to issues such as flood resilience and health access in deprived urban and rural settings \citep{burelle2021rural, li2022visualising}.

Despite significant progress in quantifying street access, we identified two key limitations in understanding and modeling road access deprivation. First, while there are well-established metrics measuring street access to buildings such as block complexity, the quality of road infrastructure is not considered by these metrics. Second, there remains a need to ground and validate such deprivation concepts through engagement with local communities, ensuring their relevance and interpretability in local contexts. Achieving this requires concepts that can be easily communicated and intuitively understood by non-experts.

To enable community-based validation of road access deprivation, this study proposes a simplified, distance-agnostic accessibility metric that captures the number of buildings obstructing the path between a structure and its nearest motorable road. We further combine this accessibility measure with road surface type to define levels of road access deprivation. In doing so, we aim to more accurately identify areas where communities are functionally disconnected from adequate road infrastructure.

The main contributions of this study are as follows:

\begin{itemize}
    \item We co-design a conceptual framework for road access deprivation tailored to urban contexts in Sub-Saharan Africa;
    \item We introduce a simple obstruction-based road accessibility metric;
    \item We generate gridded road access deprivation outputs for three cities: Nairobi (Kenya), Lagos (Nigeria), and Kano (Nigeria);
    \item We validate the model outputs using community-sourced data, providing insights into the spatial distribution and classification performance of the model.
\end{itemize}

Through this work, we aim to support efforts to map and mitigate infrastructure-related deprivation and to contribute a scalable, interpretable tool for urban planners, policymakers, and community stakeholders.

\section{Study Areas and Grid}

\subsection{Study Areas}

\begin{figure*}[!h]
    \centering
    \includegraphics[width=\linewidth]{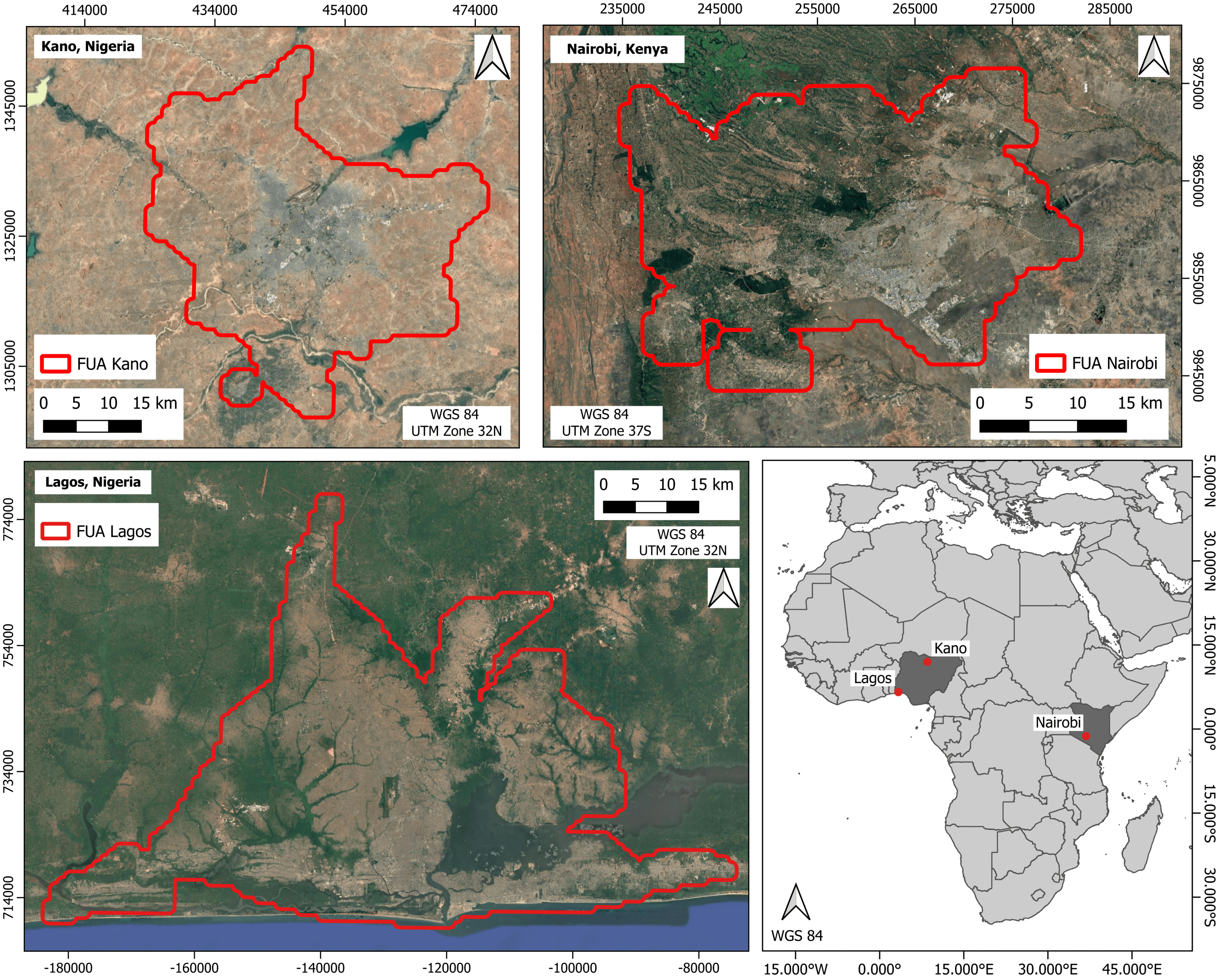}
    \caption{Overview of the study areas.}
    \label{fig:study_area}
\end{figure*}

The study areas of this research are Nairobi, Kenya, Lagos, Nigeria, and Kano, Nigeria (Figure~\ref{fig:study_area}). These cities represent different types of urbanization and geographies. Nairobi represents an inland city with very densely built-up settlements. Its administrative area spans 695 km\textsuperscript{2} with an estimated population of 4.4 million \citep{nairobi2019report}. Approximately two-thirds of the city’s urban population resides in deprived areas as of 2014 \citep{wamukoya2020nairobi}. Nairobi is also home to Kibera, often regarded as the largest informal settlement in Africa. Lagos, on the other hand, represents a coastal city with lagoon settlements and sprawling unplanned areas. It is the largest urban agglomeration in Nigeria. Different authorities estimated the population at 12.4 million and over 21 million in 2016 \citep{lagos2020context}. Kano represents a secondary city located in a conflict zone. It is also the largest city in northern Nigeria with an area of 499 km\textsuperscript{2} in 2015 \citep{yahaya2017urbanisation}. Kano state is home to an estimated population of 15 million \citep{koko2023understanding}.

To ensure that the study areas reflect the actual extent of cities and settlements, we base our analysis on functional rather than administrative extents. Administrative boundaries are often limiting since they are typically shaped by historical, political, or administrative decisions and may not necessarily reflect the actual patterns of human settlement or economic activity, especially in fast-growing African cities. For example, the administrative boundary of Nairobi has remained unchanged since 1963 and is now outdated given the substantial urban expansion of Nairobi \citep{abascal2022identifying}. Likewise, Lagos has expanded significantly beyond its city administrative boundary and is spreading to its neighboring administrative area \citep{oyalowo2022implications}. Functional urban areas (FUAs) offer a more realistic alternative, as they are defined by actual urban sprawl and human activities, encompassing the core city and economically or socially integrated surrounding regions \citep{dijkstra2019eu}. We apply a uniform buffer (1 km) to FUAs, ensuring peripheral urban zones are included.

\subsection{Grid}

Road access deprivation is modeled at a 100 m $\times$ 100 m grid. The grid system is based on the Mollweide projection system, an equal-area projection displaying the globe as an ellipse with an axes proportion of 2:1 \citep{lapaine2011mollweide}. Due to its suitability for global applications requiring accurate area representations, the grid system is also used by the global human settlement layer (GHSL) data suite \citep{pesaresi2024advances}. Consequently, this grid also facilitates an analysis of gridded population statistics, sourced from the GHSL data suite, within morphological informal areas.

\section{Methodology}

\subsection{Conceptual Model}
\label{subsec:conceptual_model}

The proposed model categorizes road access deprivation into three levels: low, medium, and high, based on road accessibility and road surface type. This level of granularity was chosen since stakeholder discussions indicated that binary distinctions between deprived and non-deprived areas fail to reflect the variation of deprivation. On the other hand, a higher level of granularity \citep[such as a degree of deprivation][]{abascal2022identifying}, can be overwhelming for non-technical stakeholders.

In the following, we provide definitions for each road access deprivation level:

\begin{definition}[\textcolor{lowc}{Low road access deprivation}]
The road network is paved and in good condition. The majority of  buildings in these areas are directly connected to motorable roads, meaning no buildings block their access to the nearest road. Areas with no buildings are also categorized as low.
\end{definition}

\begin{definition}[\textcolor{mediumc}{Medium road access deprivation}]
Roads are unpaved but motorable. Most buildings are directly connected to roads, meaning no buildings block their access to the road network.
\end{definition}

\begin{definition}[\textcolor{highc}{High road access deprivation}]
Motorable roads are sparse. Most buildings have only indirect access to the road network via paths, meaning more than one building is generally located between a building and its nearest road.
\end{definition}

\begin{figure}[p]
     \centering
     \begin{subfigure}[b]{0.8\linewidth}
         \centering
         \includegraphics[width=\textwidth]{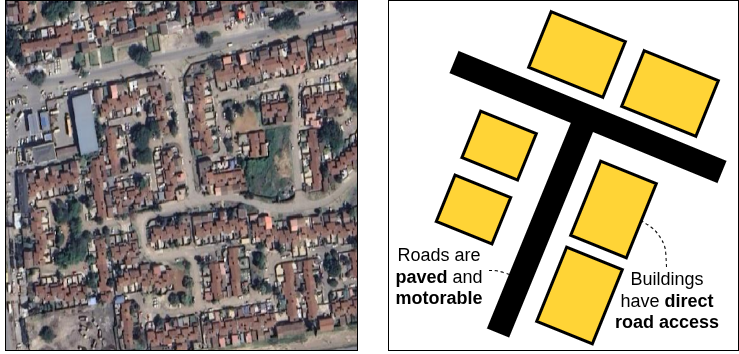}
         \caption{\textcolor{lowc}{Low road access deprivation}}
         \label{subfig:low}
     \end{subfigure}
     \hfill
     \begin{subfigure}[b]{0.8\linewidth}
         \centering
         \includegraphics[width=\textwidth]{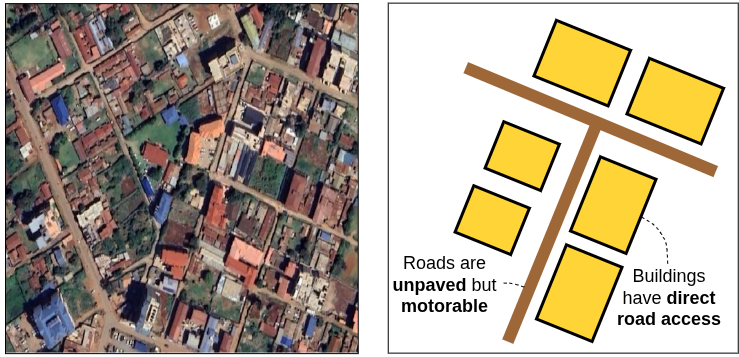}
         \caption{\textcolor{mediumc}{Medium road access deprivation}}
         \label{subfig:medium}
     \end{subfigure}
     \hfill
     \begin{subfigure}[b]{0.8\linewidth}
         \centering
         \includegraphics[width=\textwidth]{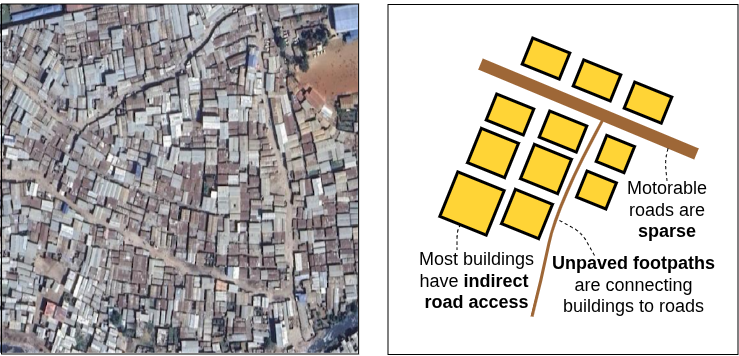}
         \caption{\textcolor{highc}{High road access deprivation}}
         \label{subfig:high}
     \end{subfigure}
        \caption{Conceptual road access deprivation model showing the definitions of the three deprivation levels: \textcolor{lowc}{low}, \textcolor{mediumc}{medium}, and \textcolor{highc}{high}. Satellite imagery: Google, Maxar Technologies.}
        \label{fig:conceptual_model}
\end{figure}

Figure~\ref{fig:conceptual_model} gives visual examples of the three road access deprivation levels using Google satellite imagery and illustrations.

\subsection{Model Implementation}

\begin{figure}[!h]
    \centering
    \includegraphics[width=\linewidth]{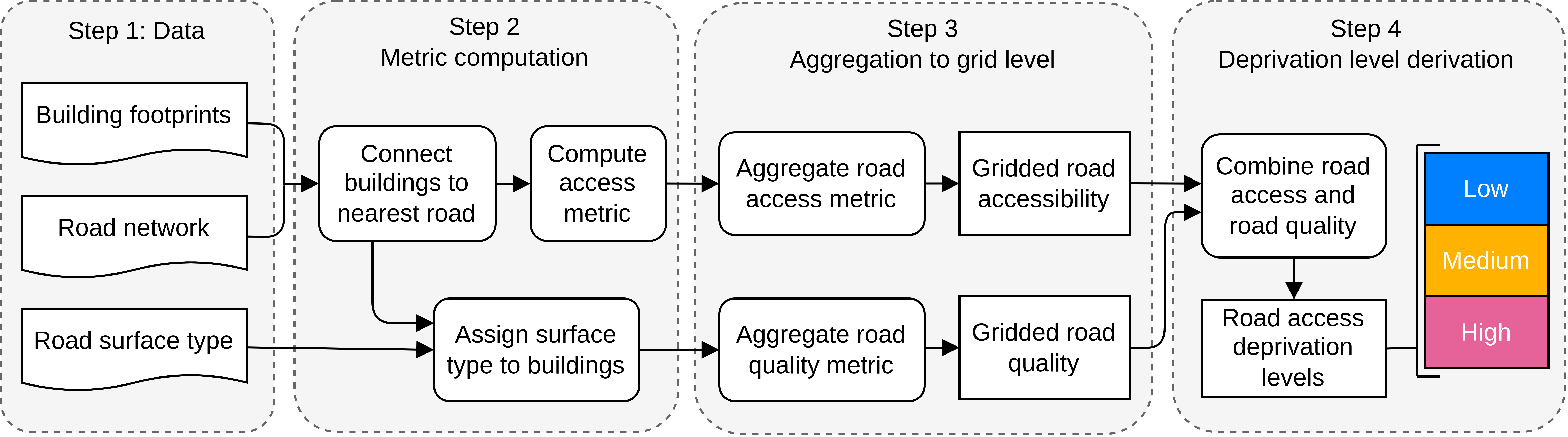}
    \caption{Overview of the road access deprivation modeling pipeline.}
    \label{fig:processing_pipepline}
\end{figure}

 The road access deprivation modeling pipeline can be summarized in four main steps (Figure~\ref{fig:processing_pipepline}). Step 1 prepares the building footprints, road network, and road surface type data. Then, a new road accessibility is computed at the building level based on the number of buildings intersected between a building centroid and its nearest road. Furthermore, information about the surface type of the nearest road is added to each building. In Step 3, these building-level metrics are aggregated to the 100 m $\times$ 100 m grid level using different mathematical operators. Finally, Step 4 applies a simple logic to classify cells into three levels of road access deprivation, low, medium, and high, based on the road accessibility metric and the predominant road surface type. The following subsections describe each step in detail. Code for the model is available from the official IDEAMAPS Data Ecosystem model repository on GitHub\footnote{\url{https://github.com/urbanbigdatacentre/ideamaps-models}}. Additionally, we created a Google Earth Engine App to visualize the input data, the metrics at the building and grid level, and the final road access deprivation levels\footnote{\url{https://ee-hafnersailing.projects.earthengine.app/view/ideamaps-road-access-deprivation}}.

\subsubsection{Data and Preprocessing}

We consider three datasets to model road access deprivation: road network, road surface type, and building footprints.

\paragraph{Road network:}

The road networks were retrieved from OpenStreetMap (OSM) \citep{haklay2008openstreetmap}. All accessible pathways and roads available in OSM are included in the road network. It should also be noted that while OSM is a crowdsourced database, its completeness for roads is generally good at a global level (above 80 \%), which is also the case for several developing countries \citep{barrington2017world}. Since we are modeling accessibility to motorable roads (see conceptual model in Section~\ref{subsec:conceptual_model}), only road classes that are considered accessible by a motorized vehicle were selected from the OSM data (see Section~\ref{subsec:road_type_selection} in Appendix).

\paragraph{Road surface type:} Road surface type data (paved or unpaved) was obtained for Kenya from \citet{zhou2024mapping}. These data were generated using a deep learning approach that determines road surface type for point locations along the OSM road network based on Google satellite imagery. The classified points were combined to determine the road surface type for a road segment. For Lagos and Kano, on the other hand, the road surface type data was obtained from \citet{zhou2025mapping}, who produced these data for Nigeria and Cameroon using a similar approach to \citet{zhou2024mapping}. However, instead of Google satellite imagery, the model was trained on 16 variables, including road network variables, socio-economic variables, and land cover variables. The accuracy assessments in both studies indicate that the resulting road surface type data are accurate (i.e., F1 scores $>$ 0.88).

\paragraph{Buildings:} The Open Buildings (V3) dataset provided by Google is used as building footprints data \citep{sirko2021continental}. This dataset is used as it has greater coverage and completeness within urban areas compared to other openly accessible datasets such as Ecopia, OSM and Microsoft \citep{chamberlain2024building}. Although Overture\footnote{\url{overturemaps.org}}, an open-data initiative that integrates several building footprint sources, was also considered for its improved completeness, issues of consistency limit its suitability for cross-city comparisons. In contrast, Open Buildings focuses on classifying buildings across Africa and much of the Global South, making it particularly appropriate for analyzing both current and emerging urban areas \citep{sirko2021continental}.

\subsubsection{Metrics Computation}

This section first introduces our new distance-agnostic accessibility metric. Specifically, we measure road accessibility based on the number of buildings intersecting lines connecting buildings to their nearest motorable roads.

We argue that this accessibility metric is preferable over distance-based metrics, as buildings in informal settlements are often not directly connected to a motorable road, yet are still near one in terms of geographic distance due to the high built-up density. Conversely, buildings in formal areas typically have direct access to motorable roads, but distances to roads can be relatively far due to larger plot sizes. Furthermore, this metric does not require footpath data, which is often incomplete or completely unavailable for informal settlements.

\begin{figure}[!h]
    \centering
    \includegraphics[width=.7\linewidth]{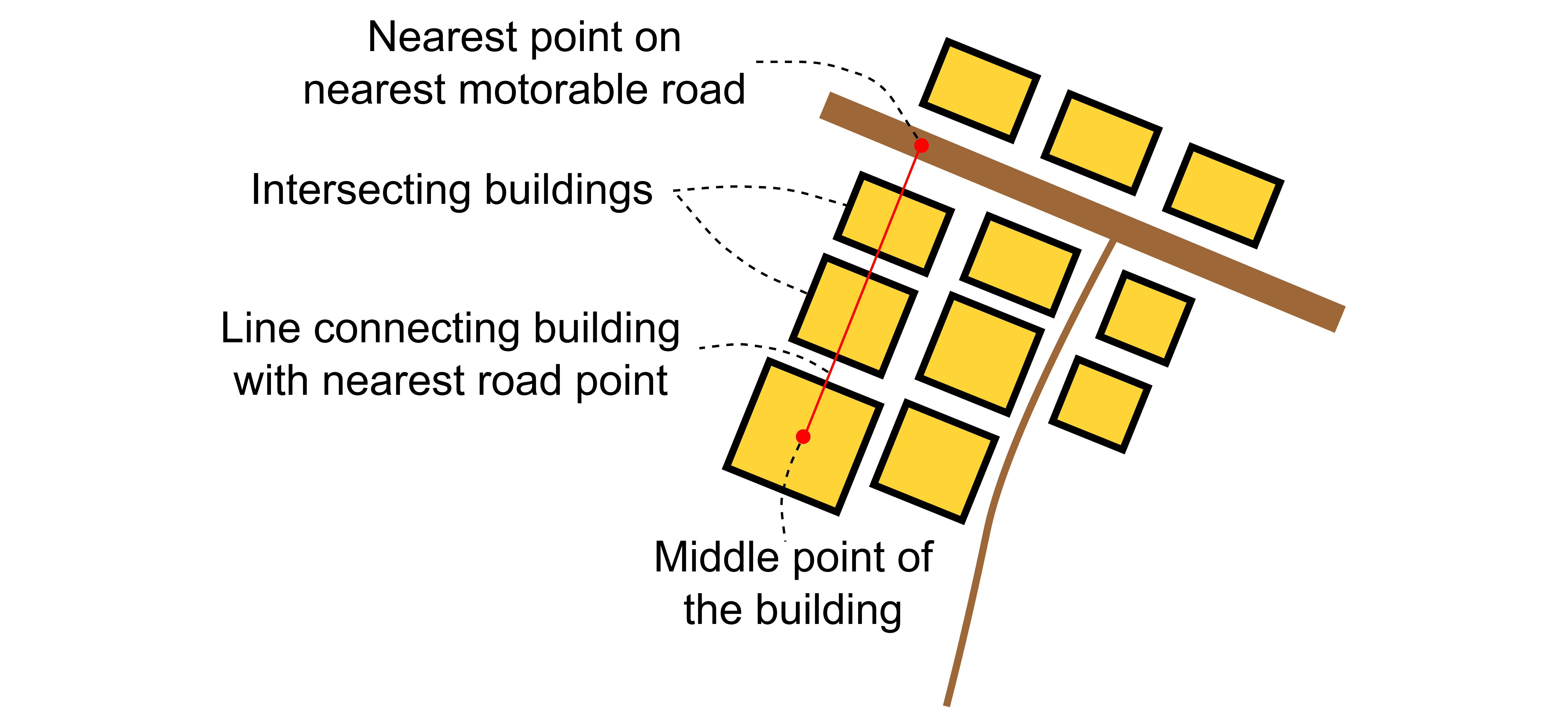}
    \caption{Illustration of the proposed distance-agnostic accessibility metric.}
    \label{fig:accessibility_metric}
\end{figure}

To compute the accessibility metric, we determine the nearest point on the nearest motorable road for all buildings. We connect this point to the middle point (i.e., centroid) of the building (Figure~\ref{fig:accessibility_metric}).

\begin{figure}[!h]
    \centering
    \includegraphics[width=.7\linewidth]{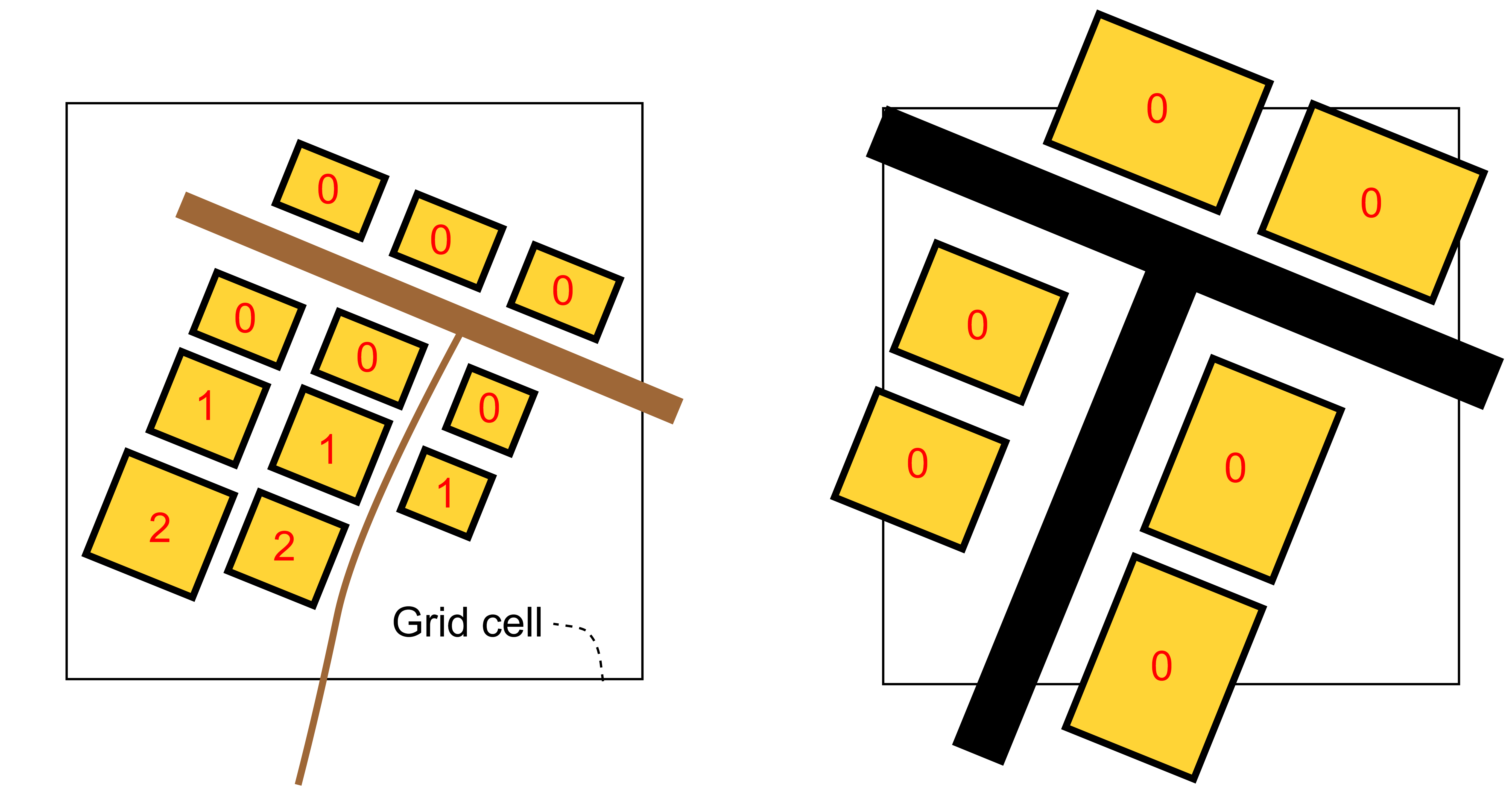}
    \caption{Examples of values for the proposed accessibility metric in (left) an informal area and (right) a formal area.}
    \label{fig:examples_accessibility}
\end{figure}

Based on the line connecting a building to its nearest motorable road point, we determine the number of buildings intersecting this line. The left and right illustrations of Figure~\ref{fig:examples_accessibility} show the resulting accessibility values for the buildings in an informal area and a formal area, respectively. Values are expected to be higher in informal areas because many buildings are only connected to motorable roads via footpaths.

In addition to the computation of accessibility values at the building level, we assign a road surface type value (paved or unpaved) to each building corresponding to the surface type of the building's nearest road. We utilize surface type as a proxy for road quality and capture whether a buildings have access to high-quality roads (i.e., paved motorable roads) or low-quality roads (i.e., unpaved motorable roads).

\subsubsection{Grid Aggregation}

Since we model deprivation at a 100 m $\times$ 100 m grid level, the building-level metrics are aggregated. For the accessibility metric, we average the values of the buildings located within a grid cell (see Figure \ref{fig:examples_accessibility} for example values that are averaged). The resulting grid-level metric represents the average number of buildings intersecting the line connecting a building to its nearest road.

For the aggregation of the road quality metric, we compute the mode of the road surface types at the building level. The aggregated metric represents the predominant road surface type that buildings within a grid cell have access to based on their nearest road.

\subsubsection{Deprivation Level Assignment}

The two metrics are combined to derive three road access deprivation levels - low, medium, and high. If the average number of intersecting buildings for a grid cell exceeds a threshold of 1, the grid cell is associated with high road access deprivation. Although direct access would require setting the threshold to 0, in practice, intersecting buildings frequently occur in formal areas due to data quality limitations, such as inaccurate building footprints data, where a single building is represented as multiple building polygons, or missing road segments.
Otherwise, if the average number of intersecting buildings is below the threshold of 1, the cell is associated with low or medium road access deprivation. The differentiation between low and medium is based on the road quality metric. Grid cells with predominant paved roads are associated with low deprivation, whereas cells where unpaved roads are predominant are associated with a medium level.

\subsection{Model evaluation}
\label{subsec:model_evaluation}

The model is evaluated based on community-sourced reference data, which is available for a limited number of grid cells. These validation data were collected using the IDEAMAPS Data Ecosystem Platform\footnote{\url{http://www.ideamapsdataecosystem.org/}}. Participatory action research sessions with two local partner communities of the IDEAMAPS Data Ecosystem project in each city were held, where community members validated grid cells by assigning a low, medium, or high deprivation level using the platform interface. Additionally, city-wide validation campaigns were conducted to collect validation data at the city level.

In detail, validation data were collected in Nairobi with community members from Korogocho and Viwandani on 4 and 5 June 2025, respectively. The city-wide validation campaign in Nairobi was conducted on 30 July 2025 (validations on 1 August 2025 were also included).  For Lagos, the city-wide validation campaign, including members from the local partner communities Okerube and Ajegunle-Ikorodu, was conducted on 11 and 12 June 2025. The inclusion of data up to 16 June 2025 accounts for substantial community validation that extended beyond the scheduled sessions. Validation data in Kano were collected for the partner communities on 22 May 2025, and the city-wide validation campaign was conducted from 27 June to 3 July 2025.

Table \ref{tab:ref_data} lists the number of people and validations per study area for road access deprivation. In total, 131 people validated 15,361 cells across the three cities. However, different people validated some cells multiple times. For these multi-validation cells, we checked for consensus among the validations. Specifically, a consensus is reached if one level receives a majority of validation votes (i.e., at least one more vote than any other level). Grid cells without a clear majority (i.e., two or more levels receive the same highest number of votes) are excluded due to a lack of consensus. Cells with only one validation are considered to have consensus by default. This was done to ensure the reliability and consistency of the ground truth, as conflicting responses could indicate uncertainty or disagreement about the level of deprivation in those locations.

\begin{table}[!h]
\caption{Overview of validation data for road access deprivation.}
\label{tab:ref_data}
\footnotesize
\begin{center}
\begin{tabular}{lrrrrrrr}
\toprule
& \# People & \# Validations &  \# Validated cells & \multicolumn{4}{c}{\# Validated cells with consensus} \\
\cmidrule{5-8}
 & & & & Total & Low & Medium & High \\
\midrule
Nairobi & 59 & 6,456 & 5,643 & 5,479 & 3,857 & 971 & 651 \\
Lagos & 52 & 5,586 & 4,478 & 4,363 & 3,418 & 555 & 390 \\
Kano & 20 & 3,319 & 3,295 & 3,280 & 1,536 & 1,104 & 640 \\
\bottomrule
\end{tabular}
\end{center}
\end{table}

Based on this reference data, model outputs are evaluated using the metrics overall accuracy (Acc) and F1 score, defined as follows:

\begin{align}
    Acc &= \frac{TP + TN}{TP + TN + FP + FN}, \\
    F1 &= \frac{TP}{TP + \frac{1}{2} \times (FP + FN)},
\end{align}

where TP, TN, FP, and FN denote the number of true positives, true negatives, false positives, and false negatives pixels, respectively.

\section{Results}

\subsection{Modeling Outputs}

\begin{figure*}[!h]
    \centering
    \includegraphics[width=\linewidth]{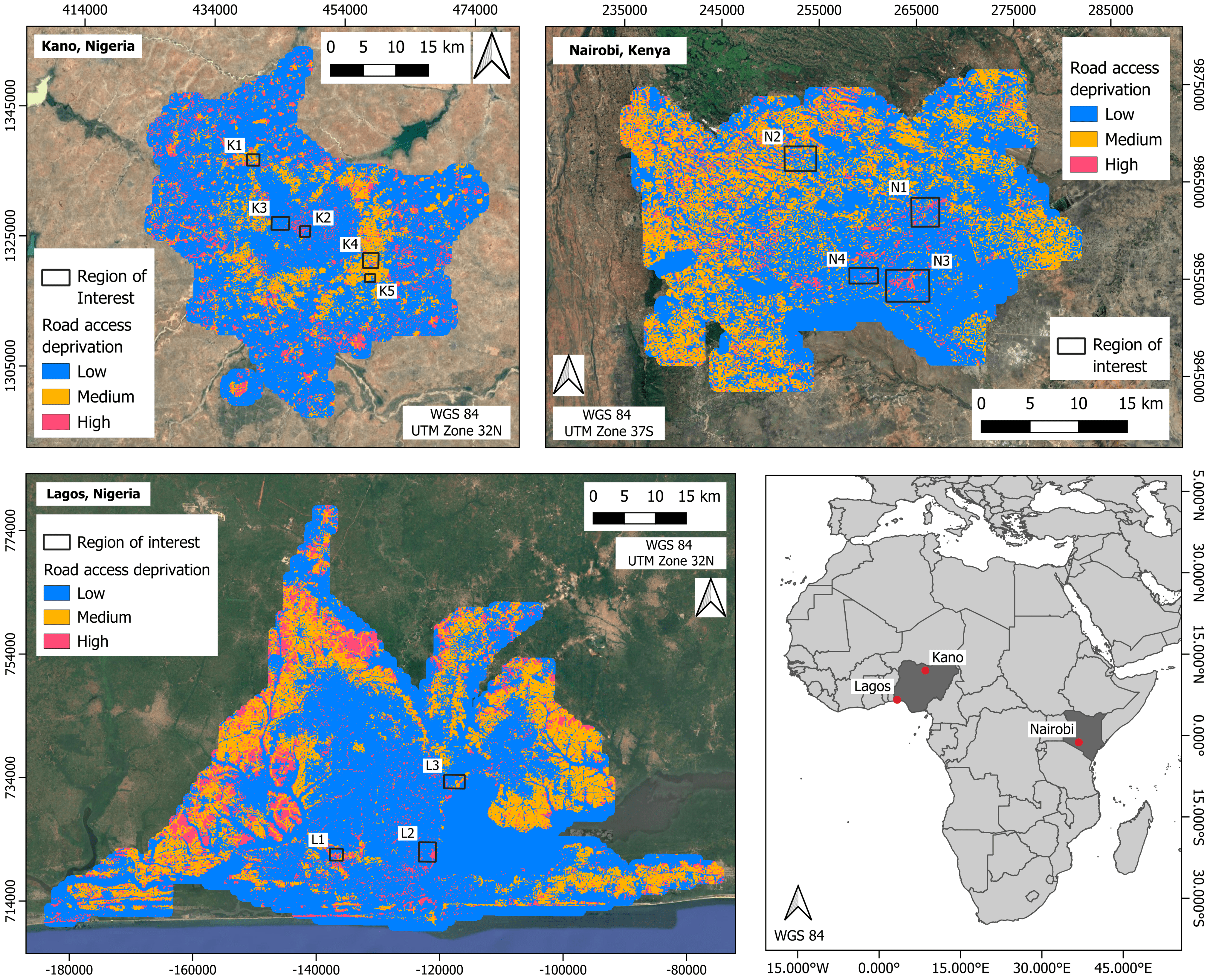}
    \caption{Road access deprivation modeling results for Nairobi, Kenya, Lagos, Nigeria, and Kano, Nigeria.}
    \label{fig:model_results}
\end{figure*}

Figure \ref{fig:model_results} presents the modeled road access deprivation for Nairobi (Kenya), Lagos (Nigeria), and Kano (Nigeria), showing low, medium, and high deprivation levels based on the distance-agnostic accessibility metric and predominant road surface type. The maps reveal consistent spatial patterns across the three cities. Low deprivation areas are predominant, and generally concentrated in the urban core, as well as on the outskirts of the cities. The latter are partly due to areas without any built-up area, which is also considered low-deprivation according to the conceptual model (Section~\ref{fig:conceptual_model}). Pockets of high deprivation are dispersed throughout the study areas. Medium deprivation zones, typically associated with unpaved but direct access to motorable roads, are found in peripheral neighborhoods outside the urban core.

\begin{figure}[!h]
     \centering
     \begin{subfigure}[b]{0.32\linewidth}
         \centering
         \includegraphics[width=\textwidth]{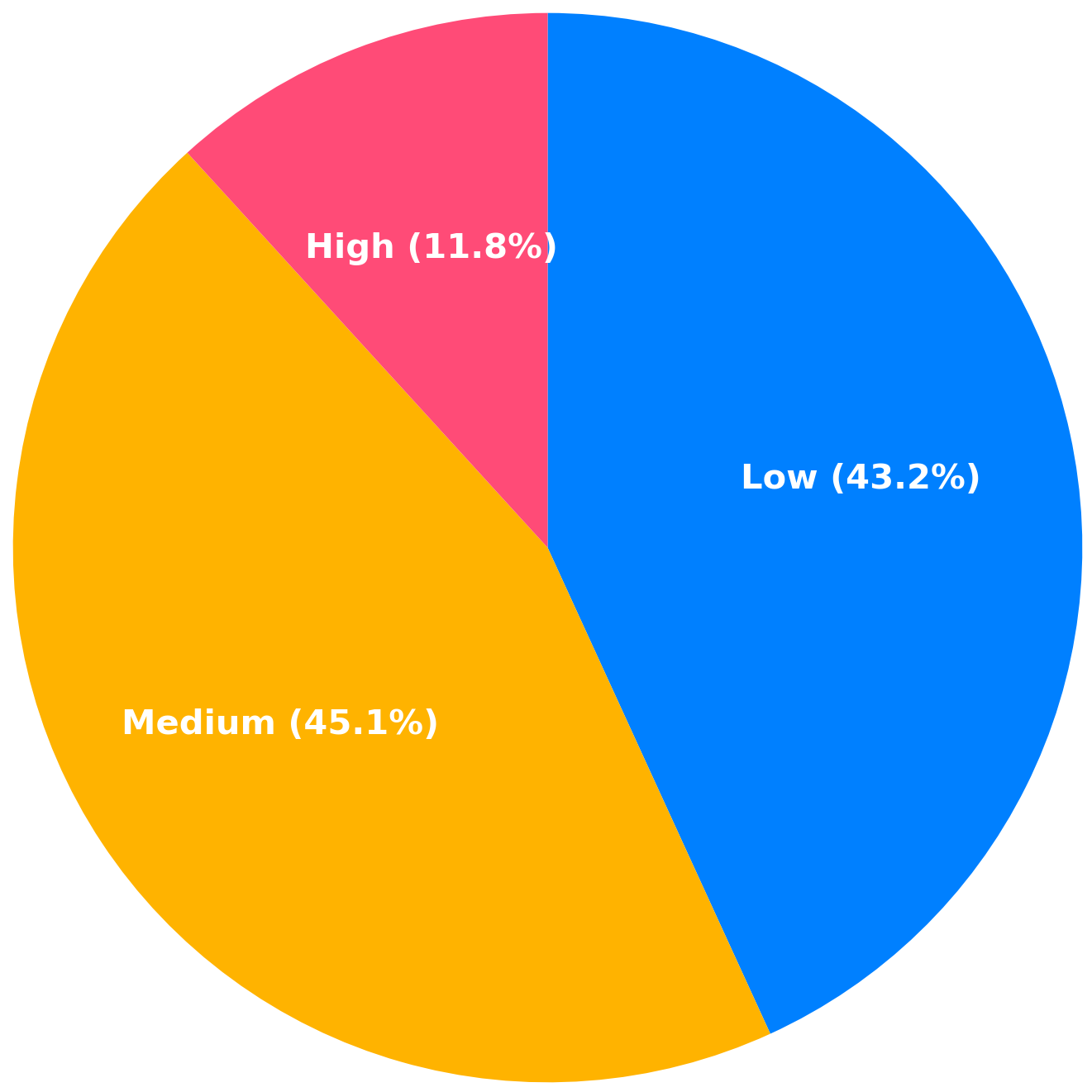}
         \caption{Nairobi}
         \label{subfig:dist_nairobi}
     \end{subfigure}
     \hfill
     \begin{subfigure}[b]{0.32\linewidth}
         \centering
         \includegraphics[width=\textwidth]{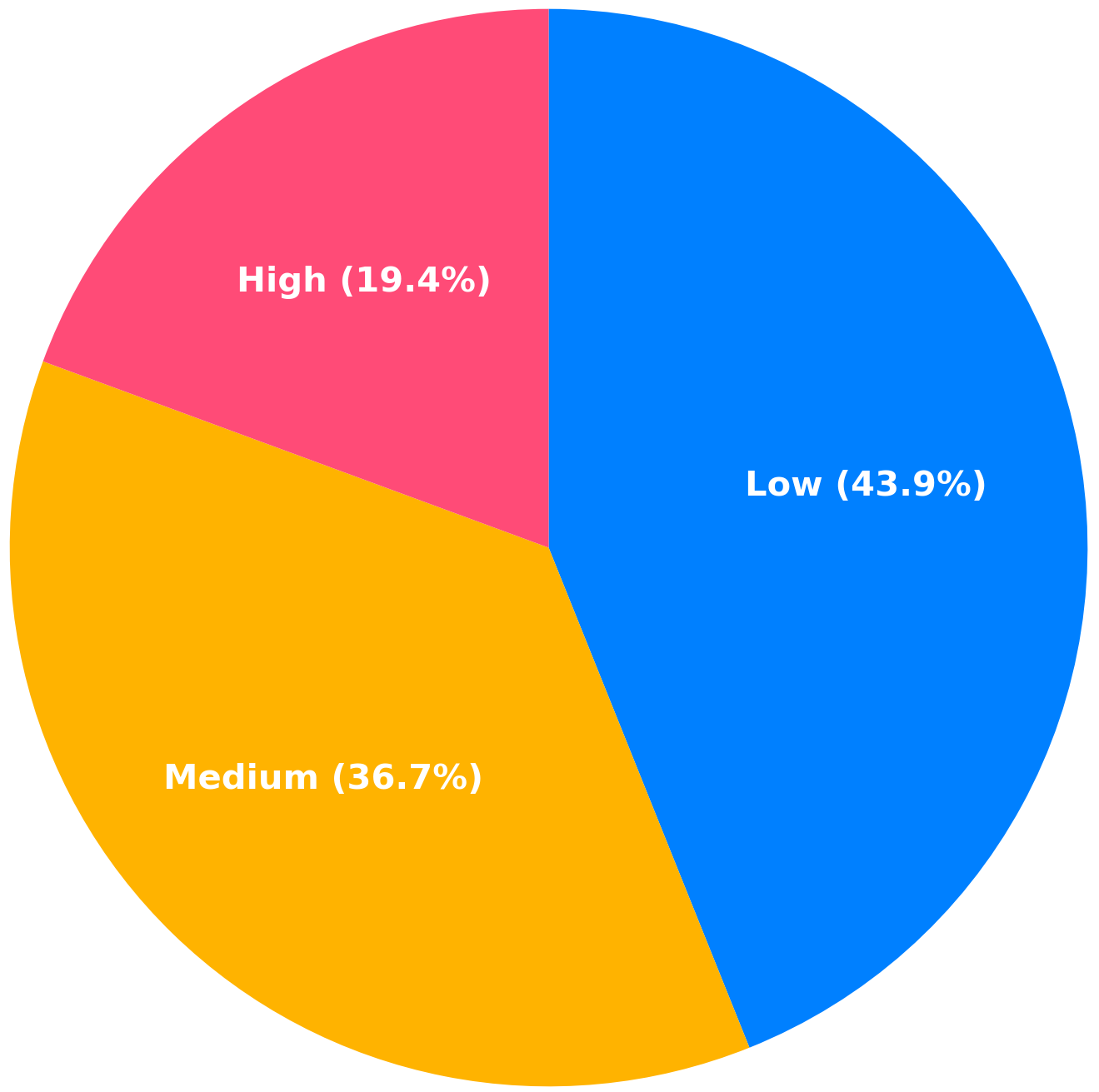}
         \caption{Lagos}
         \label{subfig:dist_lagos}
     \end{subfigure}
     \hfill
     \begin{subfigure}[b]{0.32\linewidth}
         \centering
         \includegraphics[width=\textwidth]{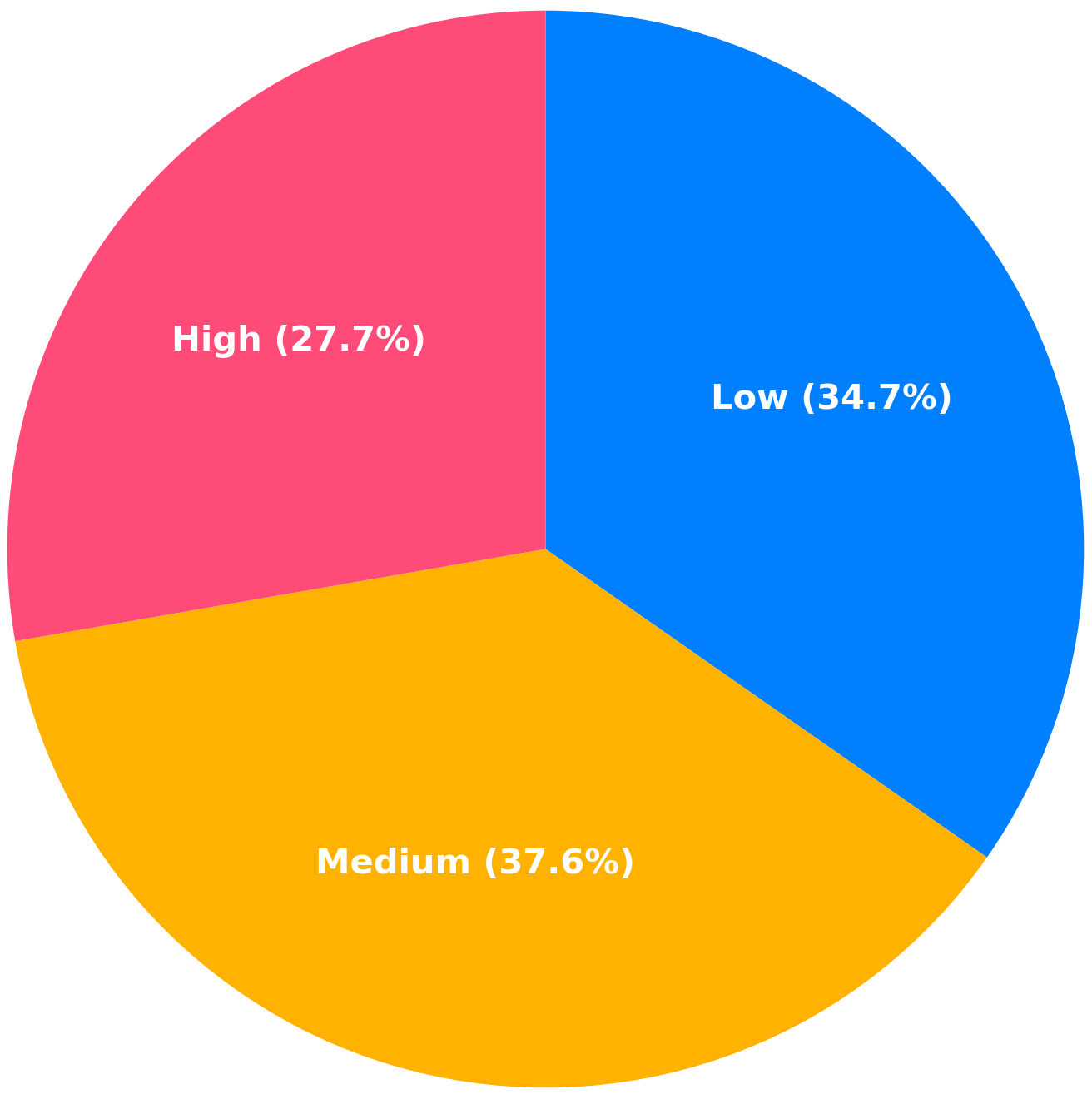}
         \caption{Kano}
         \label{subfig:dist_kano}
     \end{subfigure}
        \caption{Pie charts showing distribution of modeled road access deprivation levels, excluding cells with no buildings in the low-deprivation category.}
        \label{fig:pie_plots}
\end{figure}

In addition, Figure \ref{fig:pie_plots} shows the road access deprivation level distributions for the three cities. It should be noted that cells with no buildings were excluded to isolate road access statistics for built-up areas. In Nairobi, Lagos, and Kano, the numbers of no-building cells excluded from this analysis were 33,491 (30.3 \%), 118,471 (36.4 \%), and 99,132 (59.3 \%), respectively. In other analyses, these are included in the low-deprivation category following the conceptual model (Section~\ref{subsec:conceptual_model}). Across all three cities, low and medium deprivation areas are the most prevalent, with similar proportions of grid cells falling within either category. High-deprivation areas, on the other hand, are less common. In particular in Nairobi, relatively few grid cells fall within the high category (11.8 \%). In contrast, the percentage of highly deprived grid cells is high in Kano (27.7 \%), and only slightly lower than those for low and medium deprivation.

\subsection{Quantitative Assessment}

Table~\ref{tab:quan_assessment} presents the quantitative evaluation of the road access deprivation model against community-sourced validation data. Overall model accuracy is highest in Nairobi (79.2 \%) and Lagos (77.9 \%), followed by Kano (63.2 \%). The deprivation level-specific F1 scores indicate that the models perform best in identifying low deprivation, especially in Nairobi (F1 = 0.873) and Lagos (F1 = 0.871). Performance is poorer in identifying medium deprivation, particularly in Lagos (F1 = 0.378). For high deprivation, we observe the highest performance variance, ranging from poor results in Kano (F1 = 0.258) to good results in Nairobi (F1 = 0.691).

\begin{table}[!h]
\caption{Quantitative accuracy assessment of the road access deprivation model based on community-sourced validation data.}
\label{tab:quan_assessment}
\footnotesize
\begin{center}
\begin{tabular}{lrrrr}
\toprule
& Acc (\%) & \multicolumn{3}{c}{F1 score} \\
\cmidrule{3-5}
& & Low & Medium & High \\
\midrule
Nairobi & 79.2 & 0.873 & 0.528 & 0.691 \\
Lagos & 77.9 & 0.871 & 0.378 & 0.452 \\
Kano & 63.2 & 0.747 & 0.618 & 0.258 \\
\bottomrule
\end{tabular}
\end{center}
\end{table}

\begin{figure}[!h]
     \centering
     \begin{subfigure}[b]{0.32\linewidth}
         \centering
         \includegraphics[width=\textwidth]{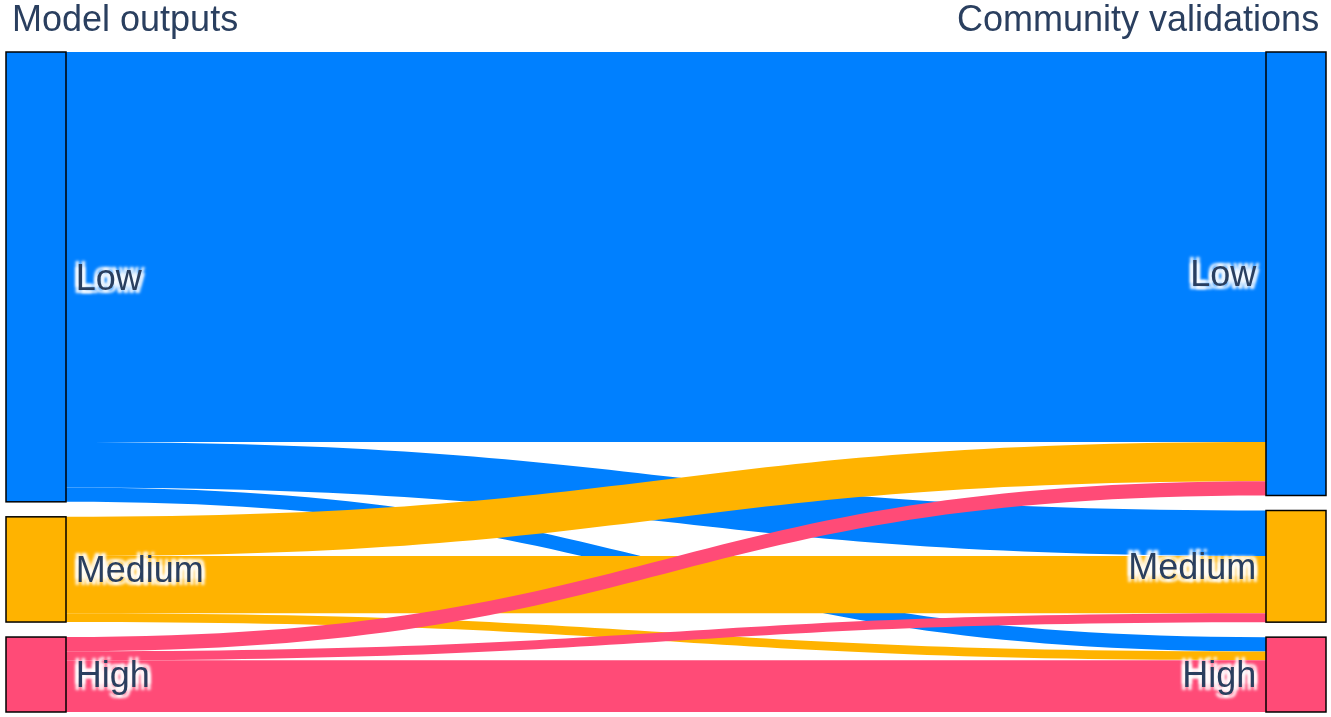}
         \caption{Nairobi}
         \label{subfig:ap_nairobi}
     \end{subfigure}
     \hfill
     \begin{subfigure}[b]{0.32\linewidth}
         \centering
         \includegraphics[width=\textwidth]{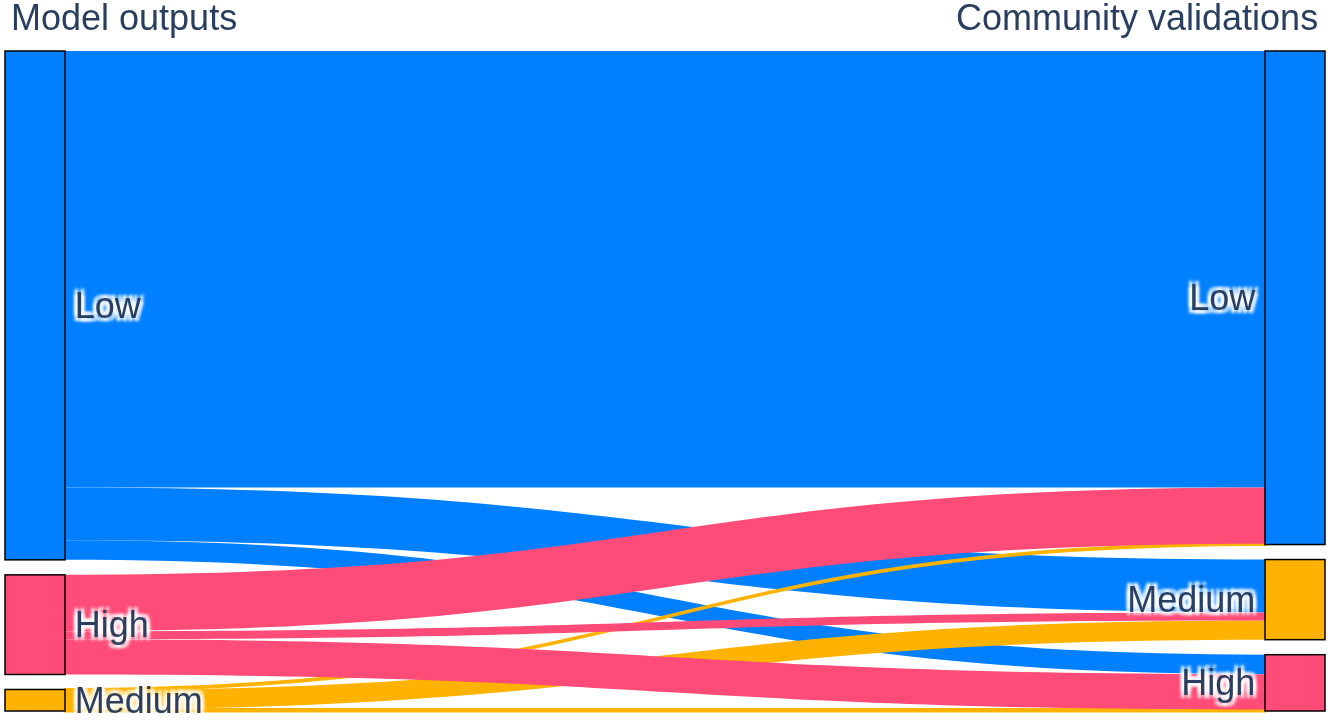}
         \caption{Lagos}
         \label{subfig:ap_lagos}
     \end{subfigure}
     \hfill
     \begin{subfigure}[b]{0.32\linewidth}
         \centering
         \includegraphics[width=\textwidth]{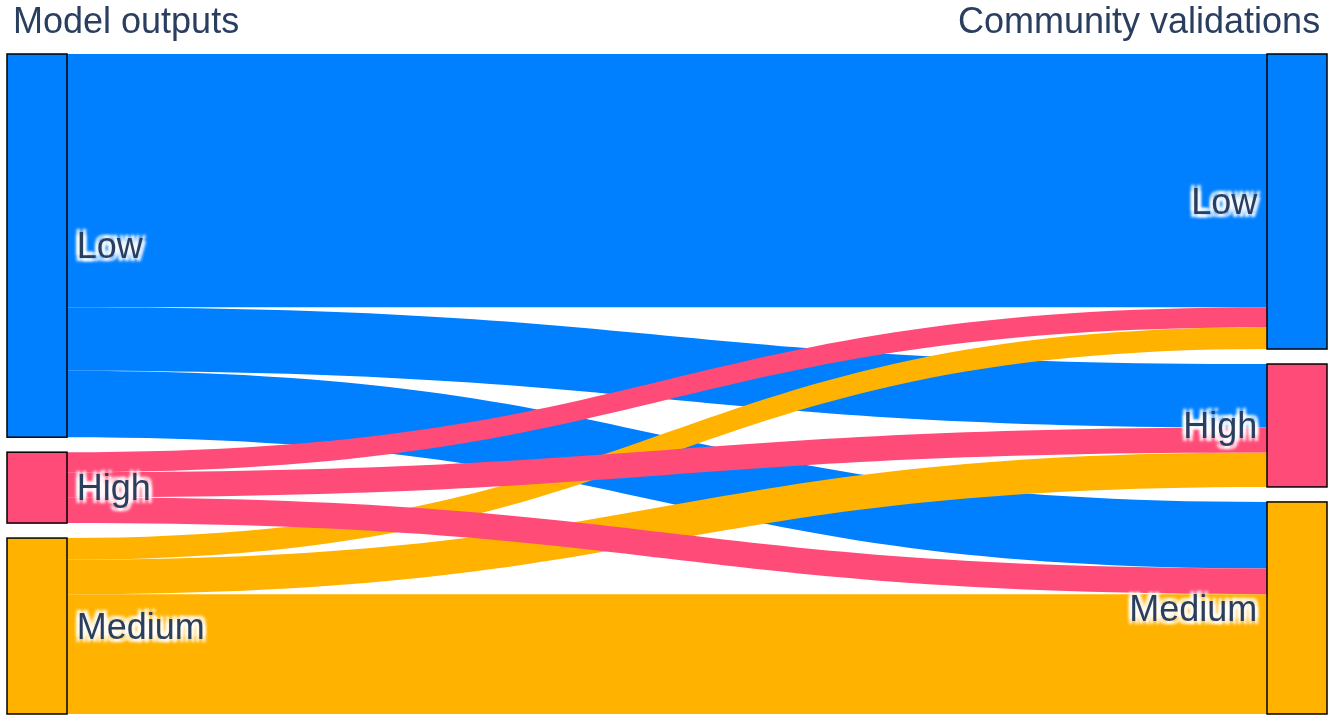}
         \caption{Kano}
         \label{subfig:ap_kano}
     \end{subfigure}
        \caption{Alluvial plots illustrating the relationship between modeled and community-validated road access deprivation levels. The left side represents modeled levels, and the right side shows the corresponding community-sourced validations. The width of each flow indicates the number of validations with a given pair of levels, highlighting areas of agreement and disagreement across deprivation levels.}
        \label{fig:alluvial_plots}
\end{figure}

Figure~\ref{fig:alluvial_plots} uses alluvial plots to visualize the relationship between model outputs and community-sourced validation data. In Nairobi and Lagos, the agreement between modeled and validated road access deprivation levels is strong for low deprivation, confirming the model's ability to identify areas with good road accessibility. Most confusion in Nairobi exists between the low and medium levels. Similarly, many of the grid cells validated as medium-deprived in Lagos were incorrectly modeled as low-deprived. We observe in Lagos that a large fraction of grid cells modeled as highly deprived were validated as low. Considerable disagreement with the modeled highly deprived grid cells is also apparent in Kano, with similar fractions of them being validated as low, medium, and high. On the other hand, the modeled representation of medium road access deprivation is most accurate in Kano.

\begin{figure}[!h]
     \centering
     \begin{subfigure}[b]{0.31\linewidth}
         \centering
         \includegraphics[width=\textwidth]{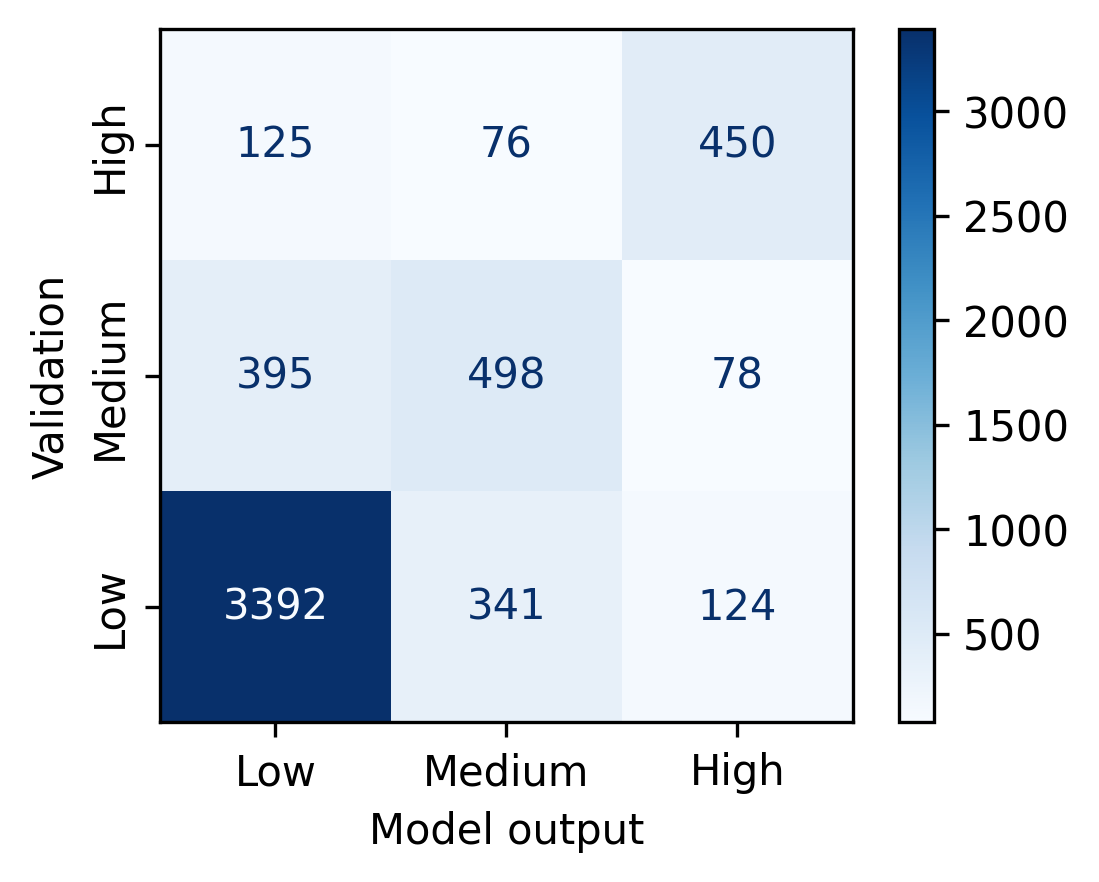}
         \caption{Nairobi}
         \label{subfig:cm_nairobi}
     \end{subfigure}
     \hfill
     \begin{subfigure}[b]{0.32\linewidth}
         \centering
         \includegraphics[width=\textwidth]{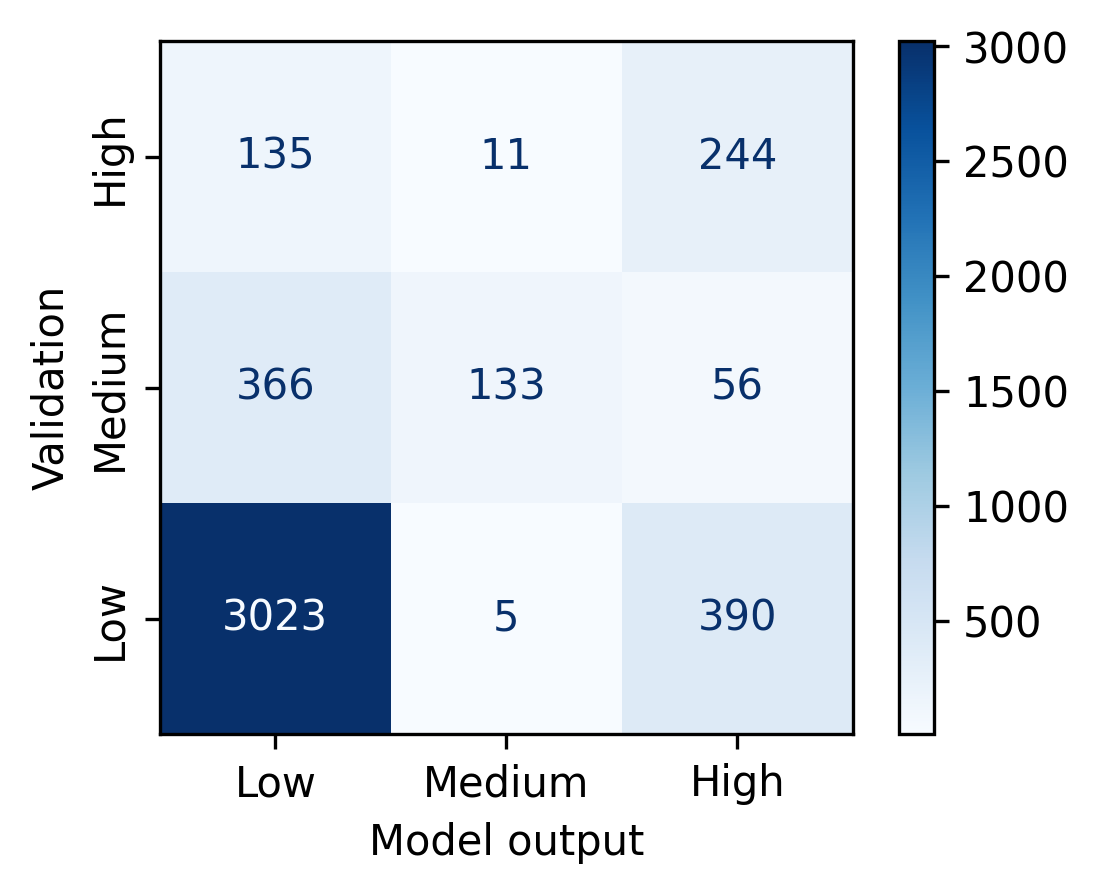}
         \caption{Lagos}
         \label{subfig:cm_lagos}
     \end{subfigure}
     \hfill
     \begin{subfigure}[b]{0.32\linewidth}
         \centering
         \includegraphics[width=\textwidth]{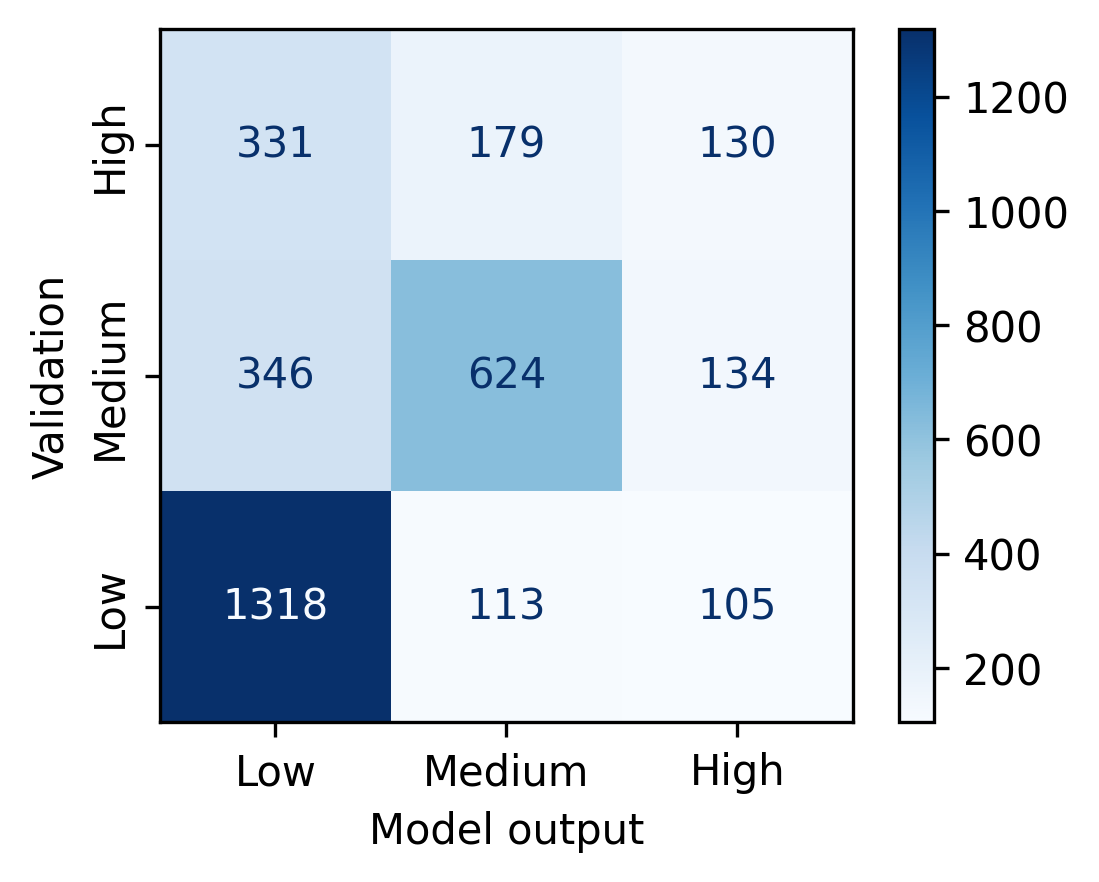}
         \caption{Kano}
         \label{subfig:cm_kano}
     \end{subfigure}
        \caption{Confusion matrix showing the road access deprivation classification performance across modeled and validated deprivation levels. Rows represent the community-sourced deprivation levels, and columns represent modeled deprivation levels. Diagonal cells indicate correct classifications, while off-diagonal cells represent misclassifications.}
        \label{fig:confusion_matrices}
\end{figure}

The confusion matrices in Figure~\ref{fig:confusion_matrices} provide a detailed view of classification performance across the three deprivation levels, complementing the accuracy metrics in Table~\ref{tab:quan_assessment} and the patterns seen in the alluvial plots (Figure~\ref{fig:alluvial_plots}). In Nairobi, the model achieves balanced performance across low and high deprivation, reflected in high F1 scores for both classes (0.873 and 0.691, respectively). The confusion matrix shows strong agreement for low deprivation, but medium deprivation is often misclassified as low, mirroring the substantial low–medium mixing seen in the alluvial plot. Lagos displays a similar pattern, with good identification of low deprivation but weaker results for medium and high deprivation. Many cells validated as medium are predicted as low, and a notable share of predicted high deprivation is validated as low, indicating overestimation of severe deprivation in some well-connected areas. Kano shows the lowest overall accuracy (63.2 \%), with high deprivation performing poorly (F1 = 0.258). The confusion matrix confirms widespread disagreement: high deprivation is almost equally redistributed across all three validated classes, while medium deprivation is captured more accurately (F1 = 0.618) than in the other cities. These patterns indicate that while the model reliably detects low deprivation in all contexts, the medium class remains the most challenging to model consistently, and high deprivation performance is highly context-dependent.

\subsection{Qualitative Assessment}

\begin{figure}[!h]
    \centering
    \includegraphics[width=.7\linewidth]{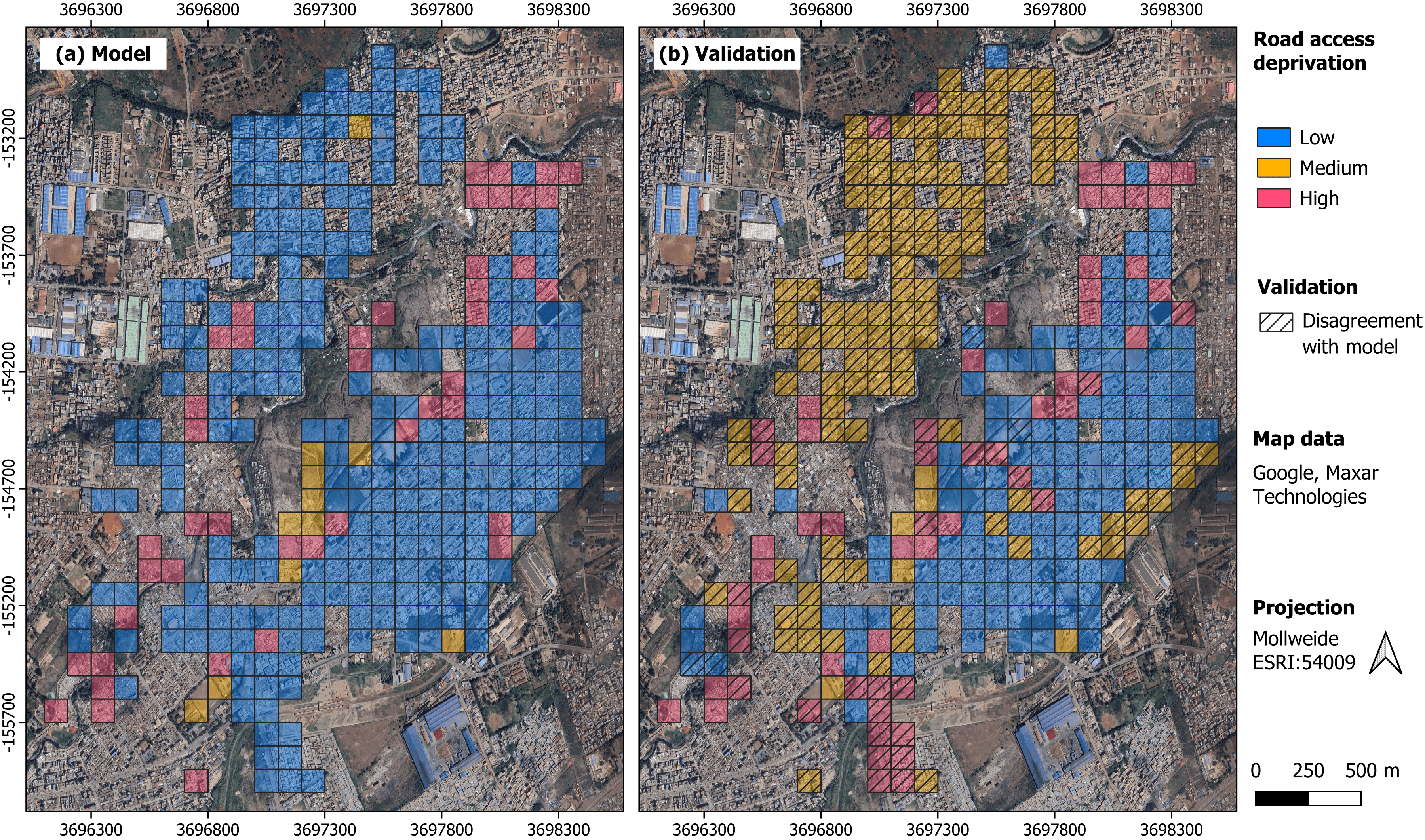}
    \caption{Region of interest N1: Visual comparison between (a) model results and (b) the data from validation with local communities in Nairobi, Kenya.}
    \label{fig:qual_assessment_n1}
\end{figure}

\begin{figure}[!h]
    \centering
    \includegraphics[width=.8\linewidth]{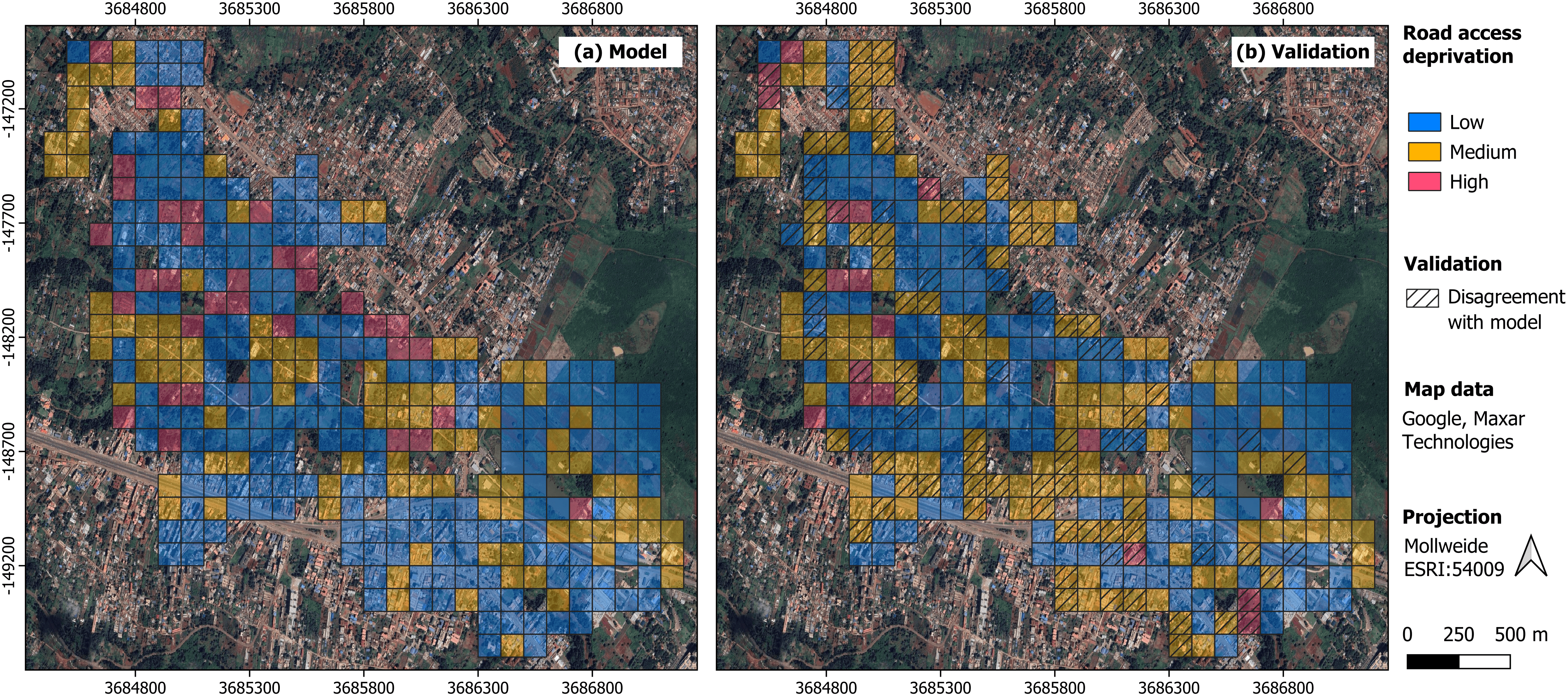}
    \caption{Region of interest N2: Visual comparison between (a) model results and (b) the data from validation with local communities in Nairobi, Kenya.}
    \label{fig:qual_assessment_n2}
\end{figure}

\begin{figure}[!h]
    \centering
    \includegraphics[width=.5\linewidth]{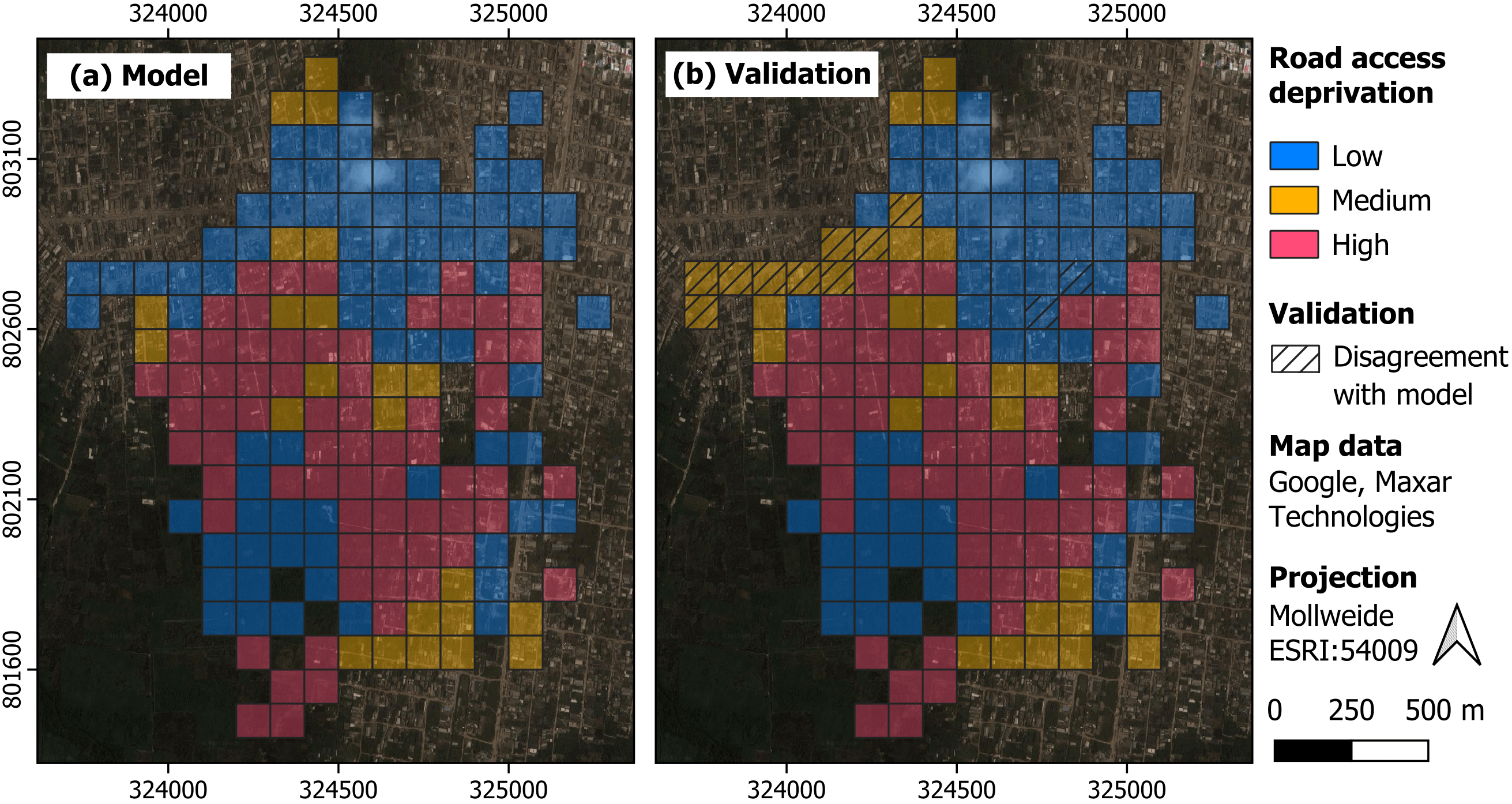}
    \caption{Region of interest L1: Visual comparison between (a) model results and (b) the data from validation with local communities in Lagos, Nigeria.}
    \label{fig:qual_assessment_l1}
\end{figure}

\begin{figure}[!h]
    \centering
    \includegraphics[width=.65\linewidth]{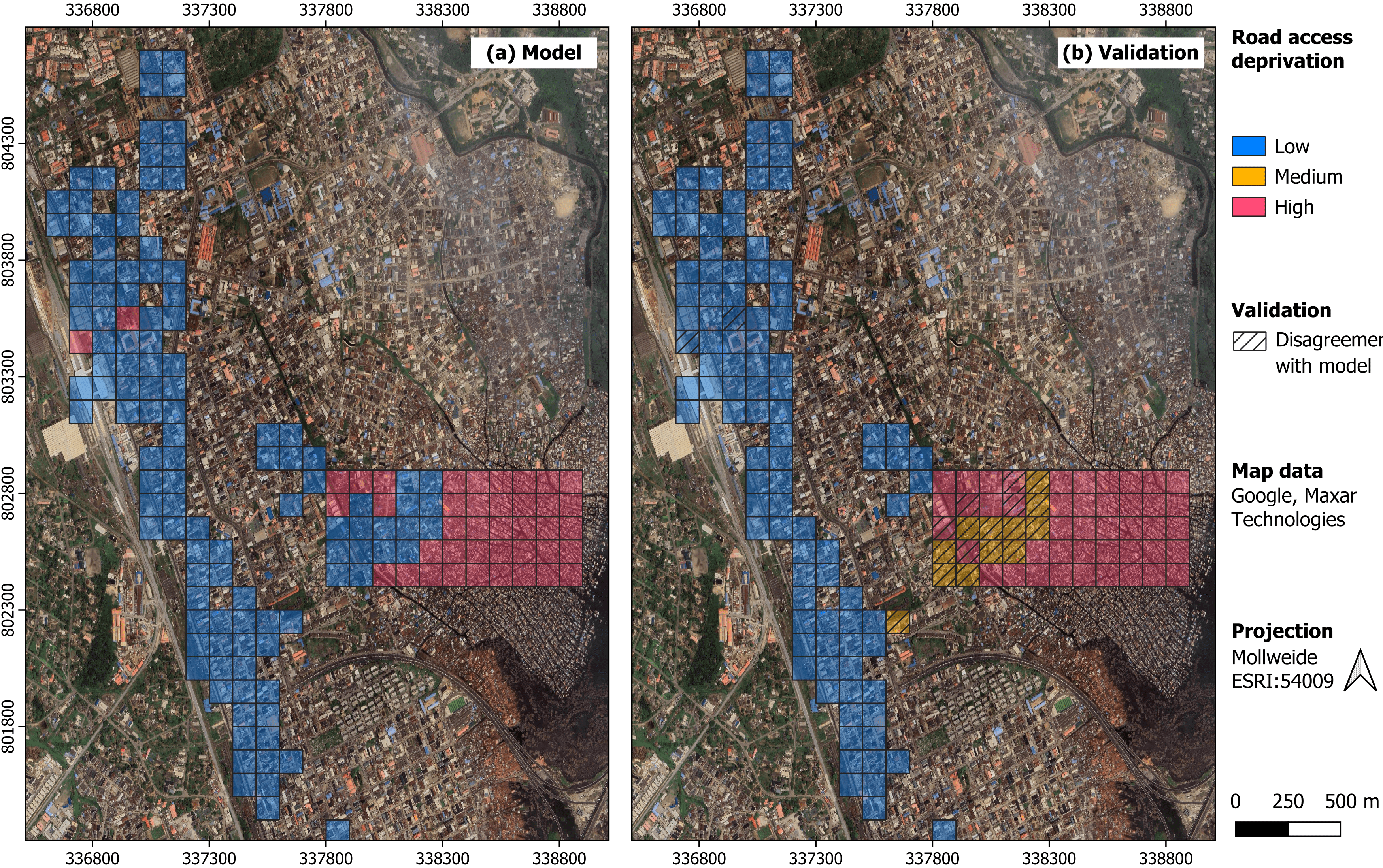}
    \caption{Region of interest L2: Visual comparison between (a) model results and (b) the data from validation with local communities in Lagos, Nigeria.}
    \label{fig:qual_assessment_l2}
\end{figure}

\begin{figure}[!h]
     \centering
     \begin{subfigure}[b]{0.5\linewidth}
         \centering
         \includegraphics[width=\textwidth]{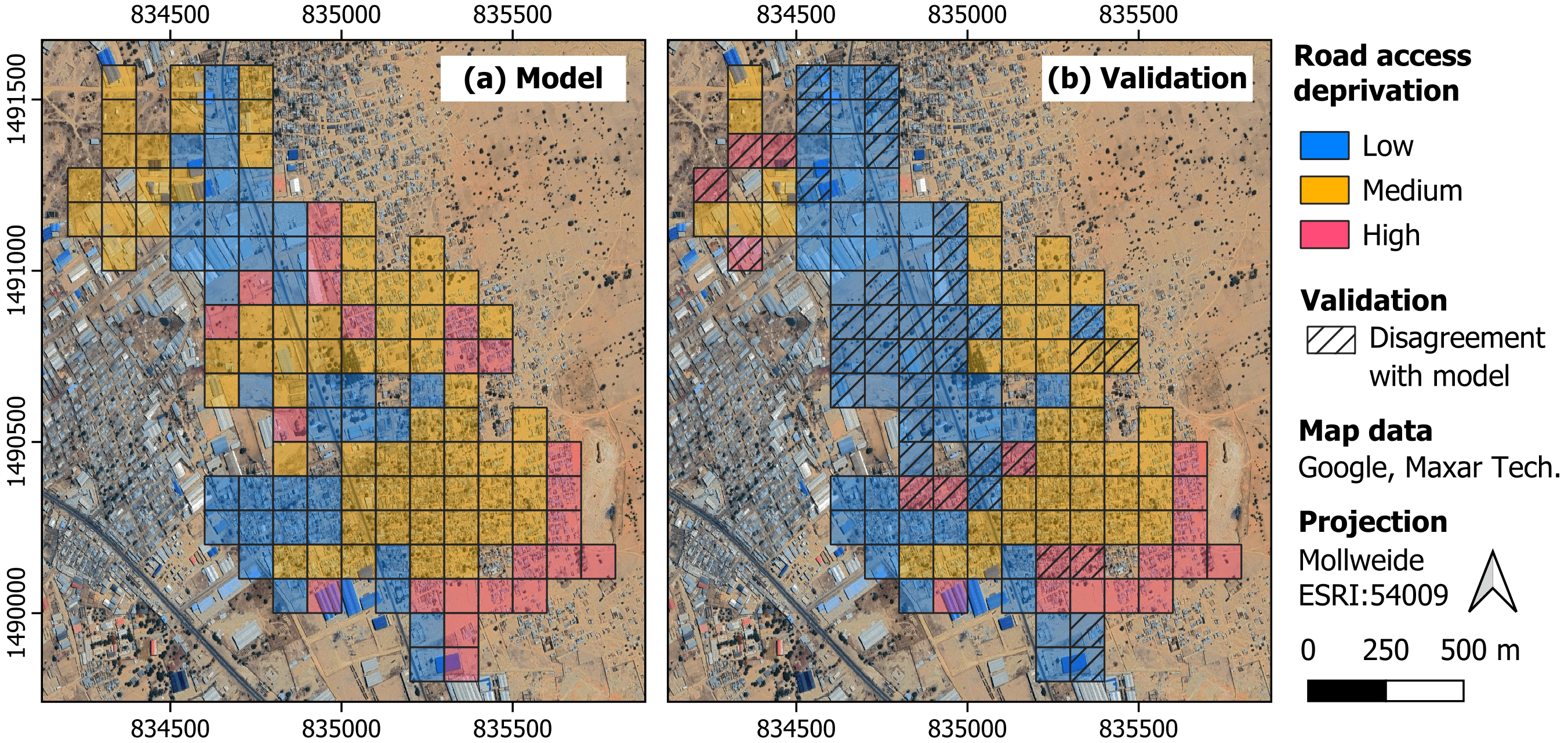}
         \caption{Region of interest K1}
         \label{subfig:k1}
     \end{subfigure}
     \hfill
     \begin{subfigure}[b]{0.45\linewidth}
         \centering
         \includegraphics[width=\textwidth]{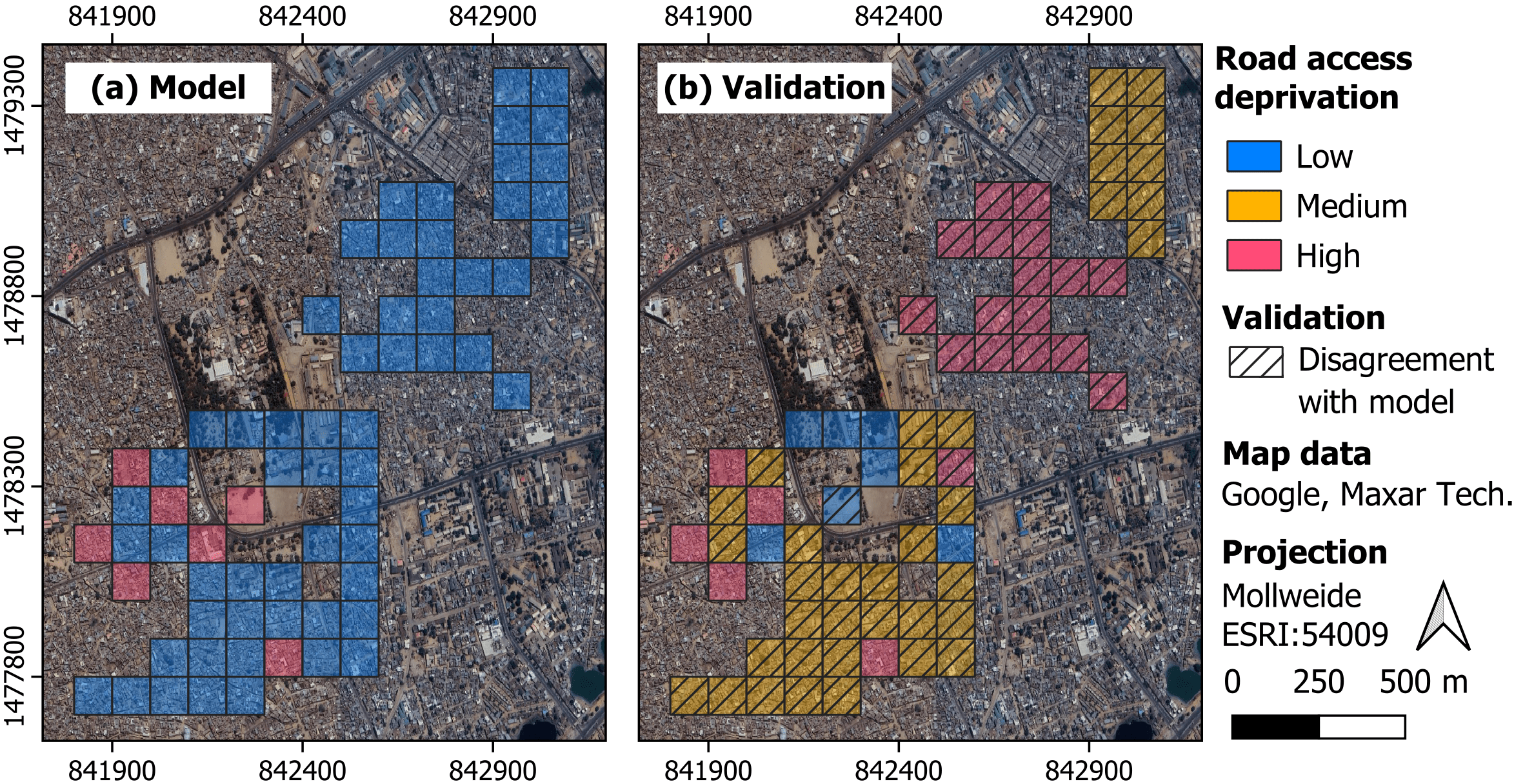}
         \caption{Region of interest K2}
         \label{subfig:k2}
     \end{subfigure}
        \caption{Visual comparison between (a) model results and (b) the data from validation with local communities in Kano, Nigeria.}
        \label{fig:qual_assessment_k}
\end{figure}

Figures \ref{fig:qual_assessment_n1} to \ref{fig:qual_assessment_k} show visual comparisons between model results and community-sourced validation data for regions of interests in Nairobi (N1: Figure \ref{fig:qual_assessment_n1}, N2: Figure \ref{fig:qual_assessment_n2}), Lagos (L1: Figure \ref{fig:qual_assessment_l1}, L2: Figure \ref{fig:qual_assessment_l2}), and Kano (K1: Figure \ref{subfig:k1}, K2: Figure \ref{subfig:k2}). Grid cells with disagreement between validations and model outputs are hashed in the validation visualization. Additional qualitative comparisons are available in the supplementary material (see Figures \ref{fig:qual_assessment_n3} to \ref{fig:qual_assessment_k5} in Section \ref{sec:supplementary_material}). Locations of all regions of interest are shown in Figure \ref{fig:model_results}.

N1 and N2, the regions of interest in Nairobi, represent a dense urban area close to the city's core and a more sparse urban area on the periphery of Nairobi, respectively. The model aligns well with the community validation data in identifying low-deprivation areas in both regions, consistent with the high F1 score for low deprivation (0.873). However, road access deprivation is underestimated in dense urban areas (N1), which is also apparent in the additional regions of interest (N3 and N4) listed in the supplementary material. Specifically, large areas validated as medium-deprived were incorrectly modeled as low, indicating that while the connectivity was correctly modeled as good, i.e., buildings are directly connected to motorable roads, the surface type of roads is predominantly unpaved instead of paved. Additionally, a few areas modeled as low were validated as highly deprived. In the rural region of interest (N2), we also observe grid cells validated as medium that were incorrectly modeled as low. In contrast to the dense urban areas, however, the model is overestimating high road access deprivation in this rural area.

L1 and L2 are both located in the periphery of Lagos. A strong agreement between modeling outputs and validation data is shown for L1, with the main confusion coming from low-deprivation model outputs that were validated as medium-deprived by community members. For L2, we also observe strong agreement for low and high road access deprivation. Disagreements in L2 primarily arise from low-deprivation areas validated as medium and from a small number of high-deprivation grid cells validated as low, the latter likely influenced by their location within clusters of low-deprivation areas.

For Kano, we show a community located in the periphery of the city (K1) and one in the urban core (K2). For K1, there is a moderate agreement between model outputs and validation data. Proximity to the primary road connecting the town to the city is validated as low road access deprivation. Medium and high road access deprivation levels are found with increasing distance to the primary road in the validation visualization. These observations are also consistent with the additional regions of interest showing peripheral neighborhoods in Kano from the supplementary material (K4 and K5).  For K2, located in the urban core of Kano, road access deprivation is modeled primarily as low, which is a severe underestimation according to the community-sourced validation data. Specifically, areas with medium and high deprivation were incorrectly modeled as low, indicating modeling errors for road surface type and road connectivity, respectively. To a much lesser degree, these errors are also present in K3 from the supplementary material, which is also located in the urban core of Kano.

\subsection{Analysis of Grid Cells with Multiple Validations}

Some grid cells received more than one validation by different community members. As described in Section \ref{subsec:model_evaluation}, we applied majority voting to determine the final deprivation level of these multi-validation grid cells, assigning the most frequently chosen level. Grid cells with no clear majority (i.e., a tie between at least two deprivation levels) were disregarded for model evaluation.

\begin{figure}[!h]
     \centering
     \begin{subfigure}[b]{0.45\linewidth}
         \centering
         \includegraphics[width=\textwidth]{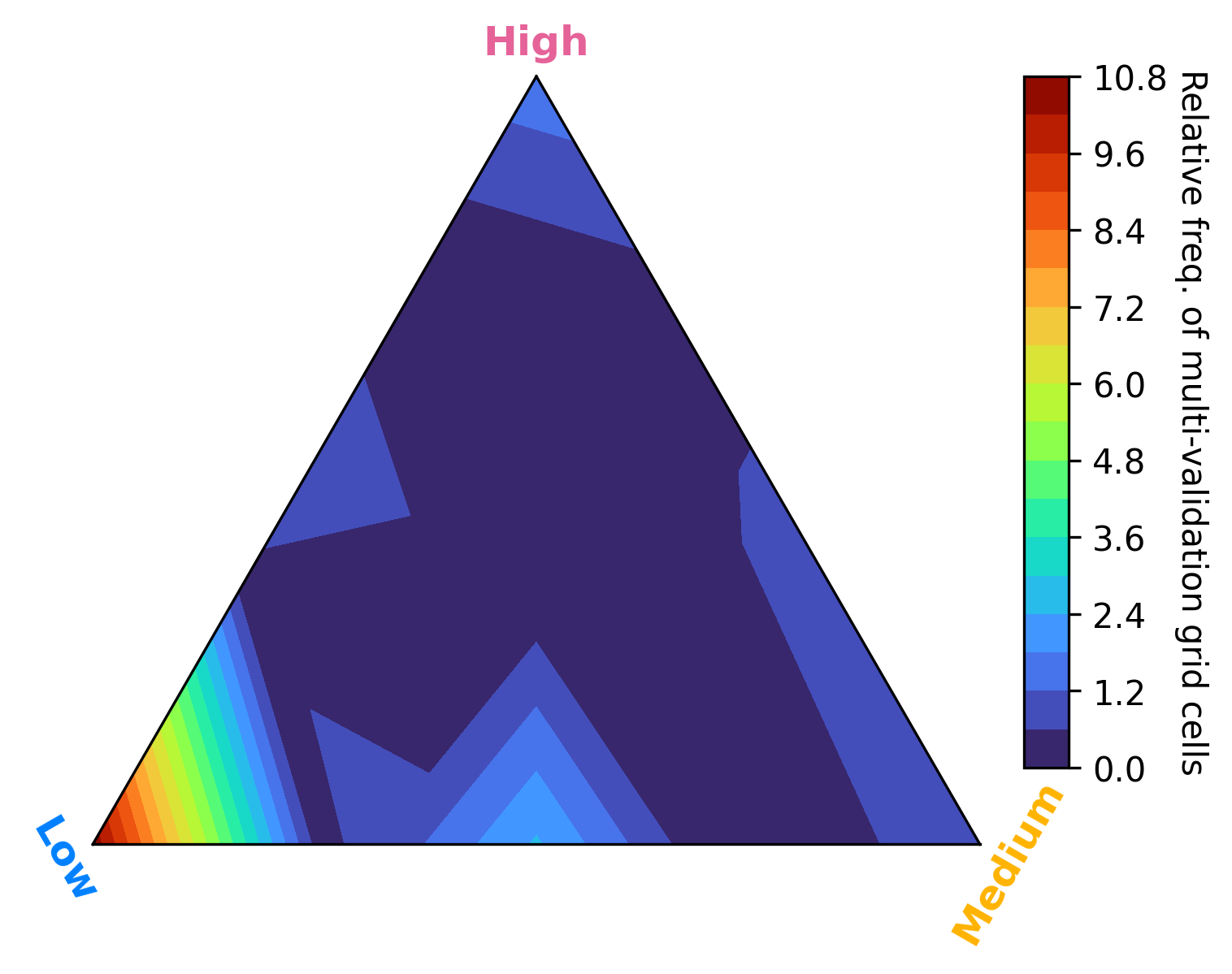}
         \caption{Nairobi ($n=721$)}
         \label{subfig:ternary_nairobi}
     \end{subfigure}
     \hfill
     \begin{subfigure}[b]{0.45\linewidth}
         \centering
         \includegraphics[width=\textwidth]{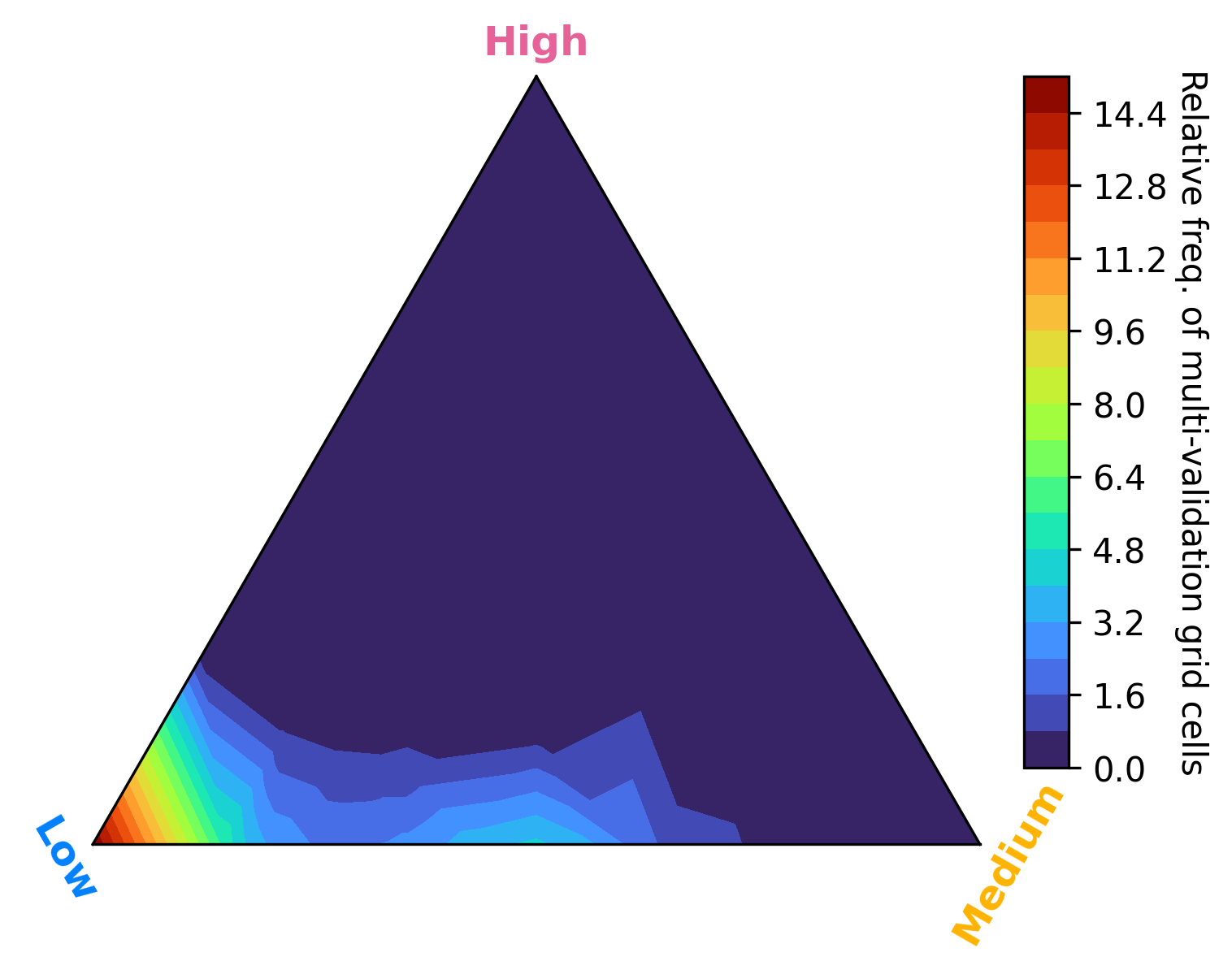}
         \caption{Lagos ($n=582$)}
         \label{subfig:ternary_lagos}
     \end{subfigure}
        \caption{Ternary plots showing the variability in validations for grid cells that were validated more than once.}
        \label{fig:ternary_par_disagreement}
\end{figure}

To better understand the extent of agreement or disagreement among participants, we analyzed the variability in assigned deprivation levels for all multi-validated grid cells (including cells with no consensus). Figure~\ref{fig:ternary_par_disagreement} shows the variability using ternary plots. Each grid cell is positioned according to the proportions of its assigned validation levels. For example, a cell validated exclusively as low (low: 1, medium: 0, and high: 0) appears in the bottom-left corner of the triangle, while a cell with equal proportions of low and medium validations (low: 0.5, medium: 0.5, and high: 0) is placed midway along the bottom axis. The color scale indicates the relative frequency of grid cells within different regions of the ternary plot.

For Nairobi (Figure~\ref{subfig:ternary_nairobi}), most multi-validated grid cells were consistently validated as low deprivation, indicating strong agreement among the community members about low road access deprivation. We also observe a few grid cells with agreement for medium and high. On the other hand, disagreement mainly exists between low and medium, as indicated by the light blue area along the axis between low and medium. To a lesser extent, disagreement is observed between medium and high, and low and high. We observe similar patterns to Nairobi in Lagos (Figure~\ref{subfig:ternary_lagos}), with a strong agreement among community members for low deprivation and some disagreement between low and medium. For Kano, the ternary plot visualization is not applicable due to the small number of multi-validation grid cells ($n=24$). For one-third of these cells, we observe agreement on low deprivation. For the other two-thirds, we observe disagreement between low and medium (3 cells) and medium and high (13 cells).

In summary, analyzing grid cells that were validated by more than one community member reveals that most often, there is agreement between validations. When there is disagreement, it typically exists between adjacent deprivation levels, meaning between low and medium or medium and high.

\FloatBarrier
\section{Discussion and Perspectives}

\subsection{Discussion}

The results indicate that the proposed distance-agnostic road accessibility metric is effective at detecting extreme cases of deprivation, i.e., well-connected and severely disconnected areas. This is particularly apparent in Nairobi and Lagos, where overall accuracies and F1 scores for low deprivation exceed 77.0 \% and 0.870, respectively. However, the consistent misclassification of medium road access deprivation suggests that modeling areas that fall between low and highly deprived areas remains challenging. This is most evident in Lagos, where nearly all medium-validated cells are predicted as low, and in Nairobi, where low–medium confusion is the dominant source of error despite comparatively balanced performance across classes.
Two potential sources may explain these shortcomings for the medium class: (1) the conceptual model for medium deprivation does not fully align with community perceptions, and (2) the operationalization of this class, particularly its dependence on road surface type, introduces classification errors.

In the current conceptual model (Figure \ref{fig:conceptual_model}), medium road access deprivation mirrors the definition of low access deprivation in terms of road connectivity, i.e., most buildings are directly connected to motorable roads, but differs in that the quality of roads is lower. To capture this, we use the predominant surface type (paved or unpaved) within a grid cell as a proxy for road quality. The low F1 scores for medium deprivation, especially in Lagos (0.378), may indicate that surface type alone is insufficient to represent the quality dimension as perceived by local communities. Furthermore, the qualitative results for towns located in the periphery of Kano indicate that with increasing distance to primary roads, road access deprivation levels are higher (Figures \ref{subfig:k1}, \ref{fig:qual_assessment_k4}, and \ref{fig:qual_assessment_k5}).

To discuss the shortcomings in the operationalization of the model, we assume that road surface type, indeed, differentiates low from medium road access deprivation. Then, the most likely source of error is the road surface type data. Specifically, we sourced automatically generated data from peer-reviewed studies \citep{zhou2024mapping, zhou2025mapping}. However, mapping road surface types is challenging, leading to confusion between paved and unpaved road segments. In turn, these errors could have affected the distinction between low and medium road access deprivation in our model. The quantitative results largely support this hypothesis, showing that many grid cells modeled as low, were validated as medium by community members (see Figure \ref{fig:alluvial_plots}). However, it should be noted that the validation data also indicates misclassifications beyond low-medium confusion.

Assuming that surface type is conceptually valid, a further explanation lies in the accuracy of the road surface type datasets used. While we sourced these from peer-reviewed studies \citep{zhou2024mapping, zhou2025mapping}, mapping paved versus unpaved segments at scale remains difficult, and misclassifications may have directly affected the low–medium distinction. The confusion matrices (Figure~\ref{fig:confusion_matrices}) and alluvial plots (Figure~\ref{fig:alluvial_plots}) show this clearly: many cells modeled as low deprivation were validated as medium. In Kano, where medium deprivation achieves a comparatively high F1 score (0.618), this distinction appears more robust, though at the expense of very low performance for high deprivation (F1 = 0.258), which is frequently predicted as low or medium.

In general, our findings on the challenge of modeling deprivation beyond the dichotomy of slums and non-slums are also reflected in the existing literature on deprivation mapping. In particular, \citet{dovey2017informal} emphasized the challenge of mapping areas that fall within strictly formal and informal categories. Furthermore, we encountered similar challenges when first developing a model to categorize areas into three levels of unplanned urbanization \citep{hafner2025towards}. However, following the iterative model improvement framework proposed in \citet{ideamaps2025ideamaps}, this baseline road access deprivation model shows great potential to be improved based on the community validations.

Beyond the challenge of modeling medium deprivation, the current model does not consider natural barriers such as rivers, steep terrain, or constructed obstructions. In reality, such barriers can significantly affect accessibility even when road networks are technically present. Their exclusion may cause the model to underestimate access challenges in specific geographies.

Finally, the community validation data, though essential, is spatially limited and potentially influenced by subjective interpretations of deprivation. This is also reflected in the observed disagreements for some grid cells within the community-sourced validation data (see Section~\ref{subsec:model_evaluation}).

\subsection{Perspectives}

This study has several potential policy and community impacts, particularly for improving infrastructure planning and supporting locally grounded decision-making. The model provides actionable insights into where buildings lack adequate connection to motorable roads, allowing planners, civil society groups, and community advocates to prioritize upgrading efforts in the most deprived areas. By visualizing deprivation spatially, the approach also helps raise awareness of how limited road access affects daily life, including emergency response, mobility, and exposure to hazards, thereby strengthening efforts toward more inclusive and resilient urban development aligned with SDG 11. At the international level, the study demonstrates how open geospatial data, interpretable modeling, and community validation can be combined to produce context-sensitive assessments of infrastructure deprivation in data-scarce regions, offering a template for wider application while underscoring the importance of citizen-generated data and the need for continued capacity building and harmonization.

The findings of this study also highlight several promising directions for future research and application. First, there is considerable potential to model road access deprivation using Earth observation data, such as high-resolution satellite imagery. For example, Earth observation datasets, such as Sentinel-2 \citep{drusch2012sentinel} or PlanetScope \citep{roy2021global}, combined with transfer learning techniques, may enable continent-scale mapping of road quality with fewer data biases. Second, refinement of the conceptual road access deprivation model logic, particularly for the medium deprivation class, may better align model outputs with community perspectives and reduce the persistent low–medium confusion. This could include integrating additional proxies for road quality (e.g., drainage, width, or surface condition) or community-defined thresholds for what constitutes adequate access. Furthermore, the incorporation of road classes according to the formal road classification scheme, distinguishing between federal road (trunk A), state road (trunk B), and local government/community road (trunk C), could improve the model. Finally, the expansion of community validation mechanisms through participatory mapping campaigns remains essential. In this study, the extension of validation campaigns beyond the initial sessions substantially increased the amount of reference data available, improving both the spatial coverage and the robustness of the evaluation. Scaling such extended campaigns in future applications could not only enhance model accuracy but also ensure that local knowledge is more comprehensively integrated into deprivation assessments.

\section{Conclusion}

This study presents a road access deprivation model that combines a distance-agnostic accessibility metric, capturing how well buildings are connected to the road network, with road surface type data as a proxy for road quality. These two components together enable the classification of urban areas into low, medium, or high deprivation levels.

The model was applied to three cities, Nairobi, Lagos, and Kano, and evaluated using community-sourced validation data. In Nairobi, the model performs well for low and high deprivation, while medium remains challenging. In Lagos, performance is also high with strong results for low deprivation but weaker performance for medium and high levels. In Kano, accuracy is lower, with poor results for high deprivation.

These results suggest that, despite variations in performance across contexts, the approach provides a solid foundation for scalable road access deprivation mapping in data-scarce regions by leveraging openly available geospatial datasets. Furthermore, future work will focus on refining the model, particularly for medium deprivation and in accounting for informal paths or natural barriers. Consequently, this model lays the groundwork for future integration into broader informality and deprivation monitoring systems, supporting efforts to prioritize underserved communities and inform targeted interventions across rapidly urbanizing regions.

\section*{Acknowledgment}
This report was generated as part of the IDEAMAPS Data Ecosystem project and the authors acknowledge the contributions from the stakeholders and partners communities in Nairobi (Kenya),  Lagos (Nigeria), and Kano (Nigeria).

The work was supported, in whole or in part, by the Bill \& Melinda Gates Foundation INV-045252. Under the grant conditions of the Foundation, a Creative Commons Attribution 4.0 Generic License has already been assigned to the Author Accepted Manuscript version that might arise from this submission. The findings and conclusions contained within are those of the authors and do not necessarily reflect positions or policies of the Bill \& Melinda Gates Foundation.

Computational resources for this work were provided by the Geospatial Computing Platform at the Center of Expertise in Big Geodata Science (CRIB), Faculty ITC, University of Twente. The authors thank the CRIB staff for maintaining the infrastructure and supporting scientific computing activities.

\FloatBarrier
\bibliographystyle{plainnat}
\bibliography{bib}


\appendix

\section{Supplemental Material}
\label{sec:supplementary_material}

\subsection{OpenStreetMap Road Type Selection}
\label{subsec:road_type_selection}

In order to model road access deprivation, we consider roads that are accessible by a motorized vehicle in our model. Therefore, we subset the OpenStreetMap (OSM) road network to motorable roads. Specifically, we use the 'class' attribute to select the following road types: 'living\_street'
, 'motorway', 'primary', 'residential', 'secondary', 'service', 'tertiary', 'trunk', 'unclassified'. Road types with class 'unknown' were also selected since the road type for many motorable roads is not specified in the data.

\FloatBarrier
\subsection{Additional Qualitative Results}
\label{subsec:additional_qualitative_results}

\begin{figure}[!h]
    \centering
    \includegraphics[width=\linewidth]{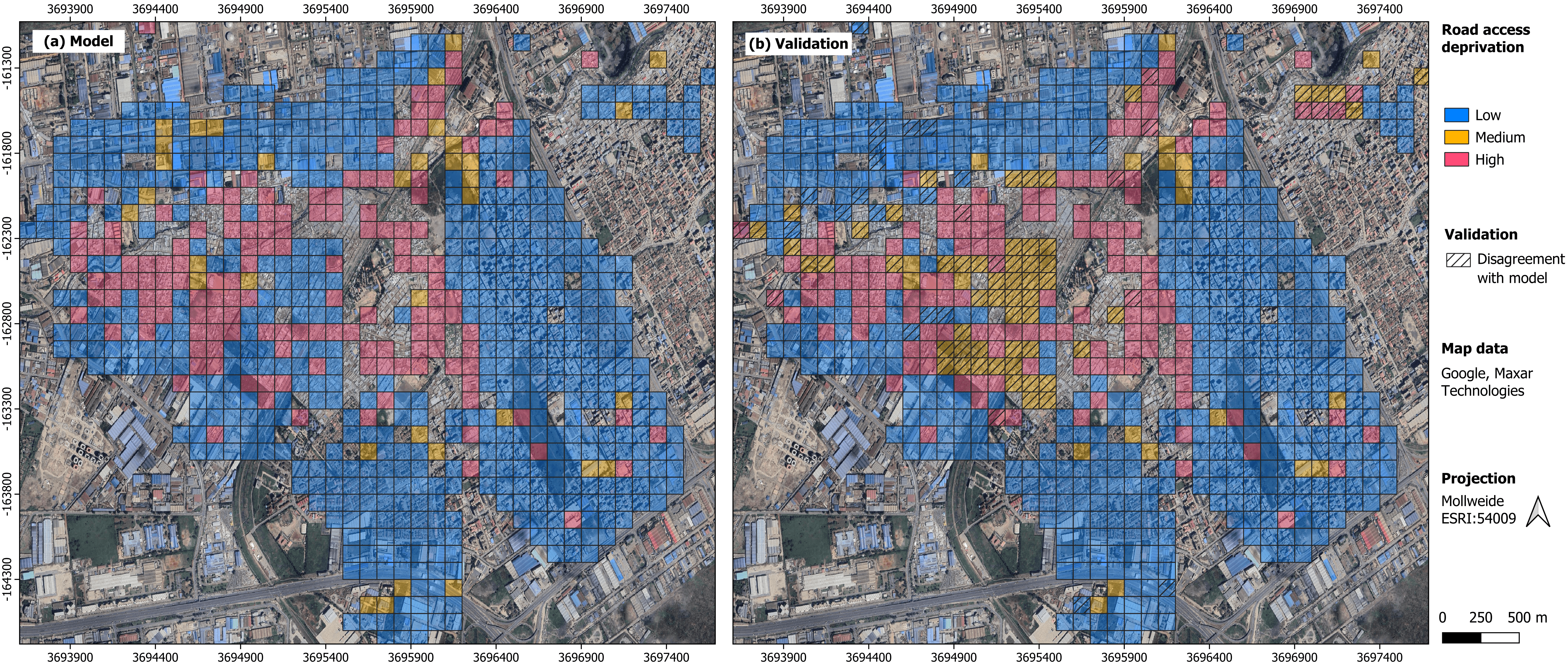}
    \caption{Region of interest N3: Visual comparison between (a) model results and (b) the data from validation with local communities in Nairobi, Kenya.}
    \label{fig:qual_assessment_n3}
\end{figure}

\begin{figure}[!h]
    \centering
    \includegraphics[width=.8\linewidth]{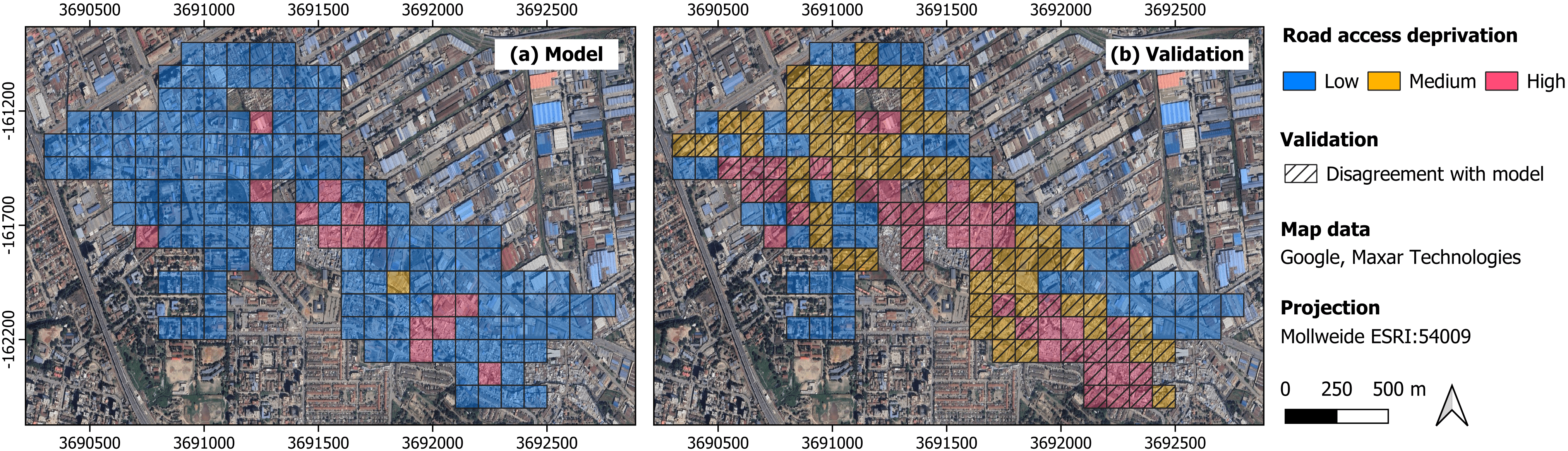}
    \caption{Region of interest N4: Visual comparison between (a) model results and (b) the data from validation with local communities in Nairobi, Kenya.}
    \label{fig:qual_assessment_n4}
\end{figure}

\begin{figure}[!h]
    \centering
    \includegraphics[width=.75\linewidth]{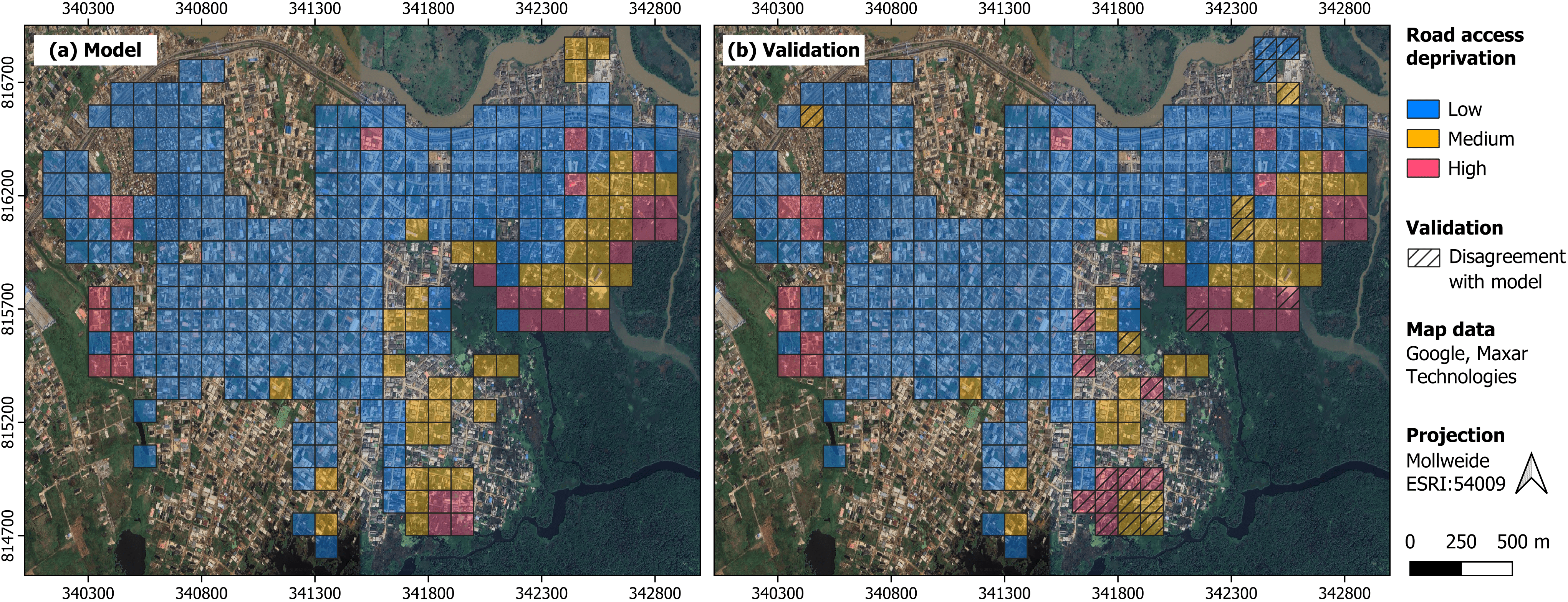}
    \caption{Region of interest L3: Visual comparison between (a) model results and (b) the data from validation with local communities in Lagos, Nigeria.}
    \label{fig:qual_assessment_l3}
\end{figure}

\begin{figure}[!h]
    \centering
    \includegraphics[width=.65\linewidth]{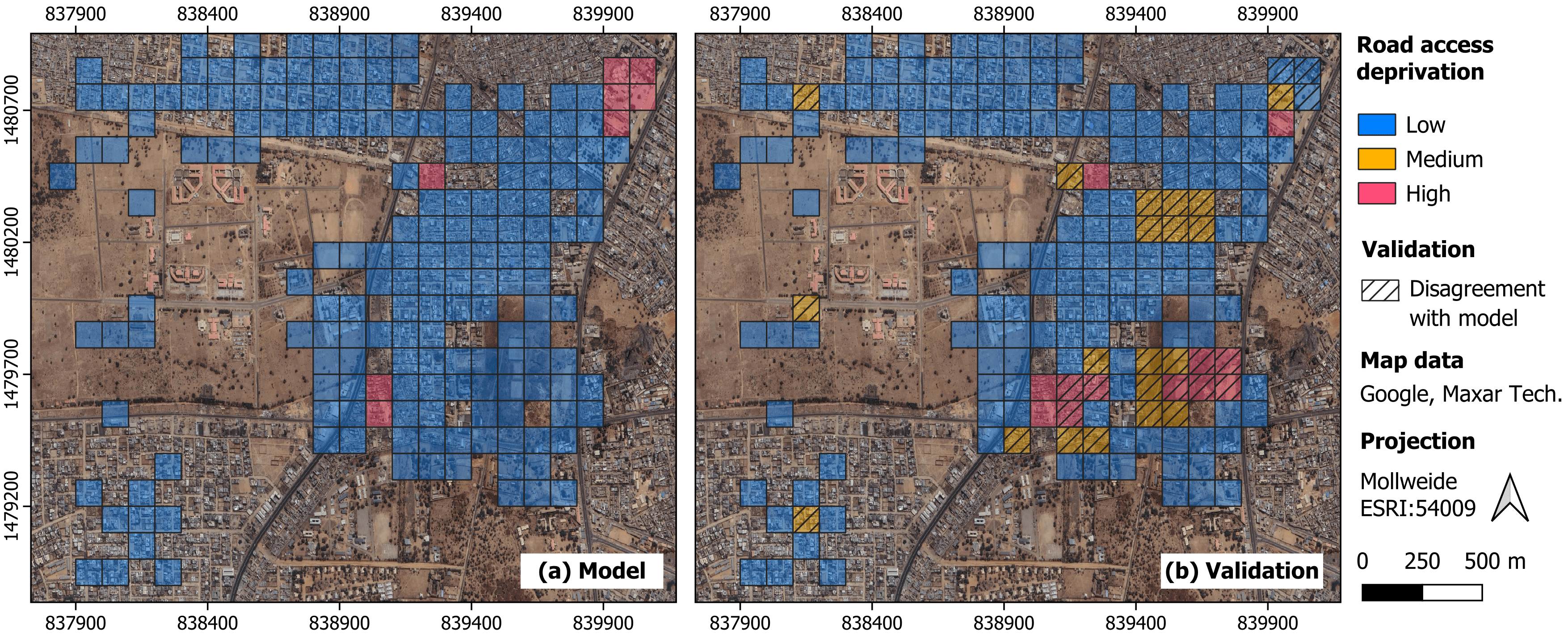}
    \caption{Region of interest K3: Visual comparison between (a) model results and (b) the data from validation with local communities in Kano, Nigeria.}
    \label{fig:qual_assessment_k3}
\end{figure}

\begin{figure}[!h]
    \centering
    \includegraphics[width=.6\linewidth]{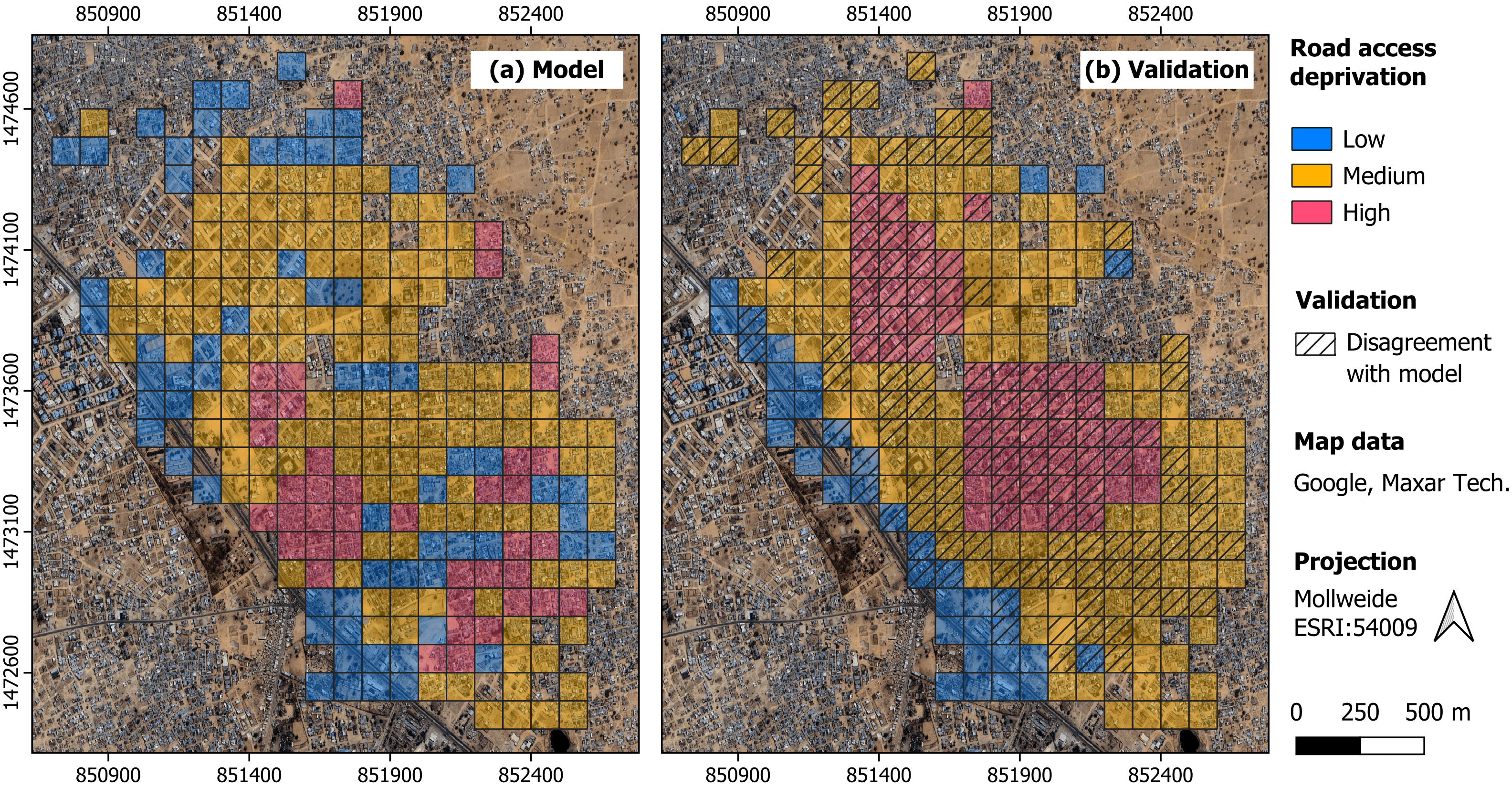}
    \caption{Region of interest K4: Visual comparison between (a) model results and (b) the data from validation with local communities in Kano, Nigeria.}
    \label{fig:qual_assessment_k4}
\end{figure}

\begin{figure}[!h]
    \centering
    \includegraphics[width=.5\linewidth]{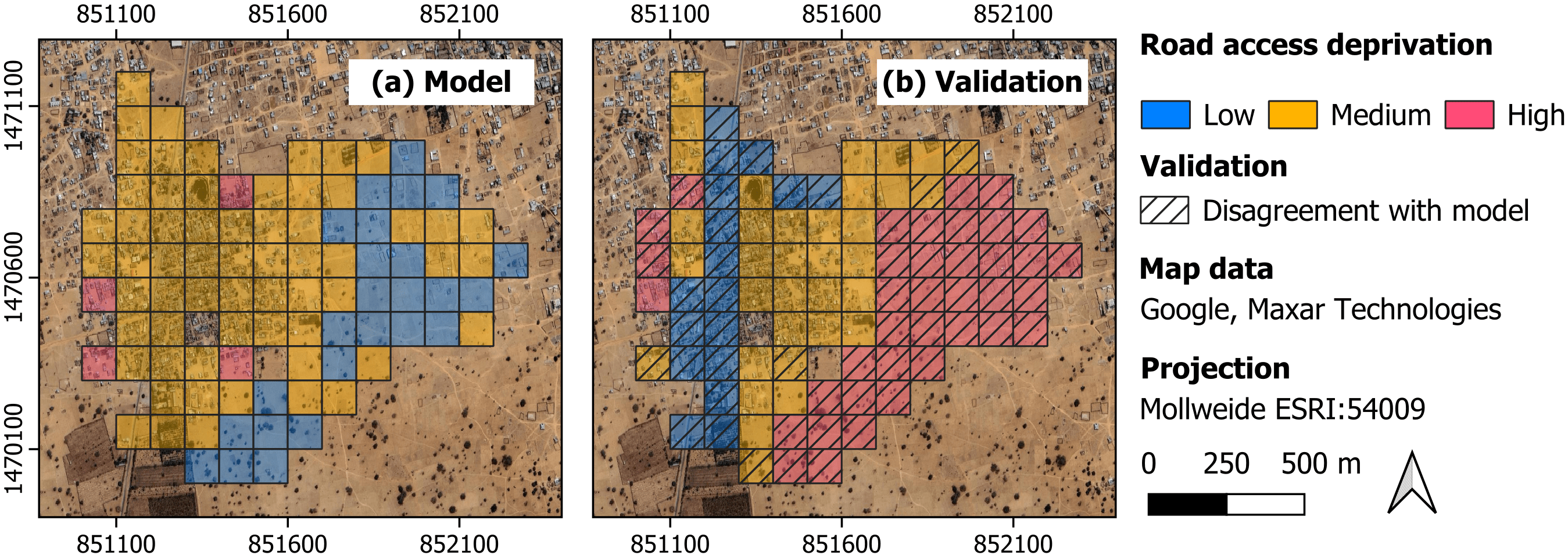}
    \caption{Region of interest K5: Visual comparison between (a) model results and (b) the data from validation with local communities in Kano, Nigeria.}
    \label{fig:qual_assessment_k5}
\end{figure}

\end{document}